\pdfoutput=1
\documentclass[cernpreprint, atlasdraft=falese, UKenglish, texmf, orcidlogo]{atlasdoc}
\usepackage[subcaption, rotating, longtable, multirow]{atlaspackage}
\usepackage[italic]{hepnicenames}
\usepackage{atlasbiblatex}

\usepackage[hepparticle, hepprocess, jetetmiss, snippets, xref=false, unit=false]{atlasphysics}
\graphicspath{{logos/}{figures_final}}

\usepackage{ANA-HIGP-2026-05-PAPER-defs}

\AtlasTitle{Search for \textit{tH} production in the \(H \to \tau\tau\) decay mode,
and a combination with other searches,
using \textit{pp} collisions at 13 TeV and 13.6 TeV with the ATLAS detector}
\AtlasAbstract{%
A search for the production of a Higgs boson in association with a single
top quark in multi-lepton final states including $\tau$-leptons is presented,
using an integrated luminosity of \lumi of proton--proton  collision data at
a centre-of-mass energy of \(\sqrt{s} = 13\,\text{TeV}\) collected by the ATLAS detector at the LHC\@.
The analysis targets the \(H \to \tau\tau\) decay,
where at least one \(\tau\)-lepton decays hadronically, to indicate the presence of the Higgs boson.
No significant excess over the expected background is observed, and an observed\,(expected) upper limit
at the 95\% confidence level on the signal strength
is set at \limtHtautauObs\,(\limtHtautauExp) times the Standard Model prediction.

This measurement is combined with previous publications of the ATLAS Collaboration
targeting the same production process but in different final states,
using \lumi of proton--proton collision data at \(\sqrt{s} = 13\,\text{TeV}\) and \lumiRunThree at \(13.6\,\text{TeV}\).
The combined measurement of the signal strength yields \(\mu_{tH} = \mutHMLbbggMinosTot = \mutHMLbbggMinos\) times
the Standard Model expectation,
corresponding to an observed (expected) \tH significance of \sigtHMLbbggObs\,(\sigtHMLbbggExp) standard deviations.
The combined observed\,(expected) upper limit at the \qty{95}{\%} confidence level on the signal strength
is set at \limtHMLbbggObs\,(\limtHMLbbggExp) times the Standard Model prediction.
This combination represents the most precise measurement of \tH production to date.
}

\AtlasRefCode{HIGP-2026-05}
\PreprintIdNumber{CERN-EP-2026-224}

\AtlasJournal{JHEP}
\hypersetup{pdftitle={ATLAS document},pdfauthor={The ATLAS Collaboration}}

\begin{document}

\maketitle

\tableofcontents
%

%
%

%
%
%
%
%
%
%
%

%
%

%

%
\section{Introduction}%
\label{sec:intro}

The Yukawa coupling of the Higgs boson to the top quark (\(y_{{\Pqt}}\)) has a value close to unity,
which implies that it plays an important role in radiative corrections to the Higgs potential~\cite{Degrassi:2012ry,Buttazzo:2013uya},
influencing the shape of the potential and, consequently, the stability of the electroweak vacuum~\cite{Englert_2014}.
The Higgs--top Yukawa coupling can be directly probed by measuring the processes
where the Higgs boson is produced in association with a top quark--antiquark pair (\ttH), or with a single top quark (\tH).
The \ttH production has been observed by both ATLAS and CMS Collaborations~\cite{HIGG-2018-13,CMS-HIG-17-035},
while experimental evidence for \tH production, which is uniquely sensitive to the sign of the top-quark Yukawa coupling, has not yet been established.

\begin{figure}[htbp]
\centering
\subcaptionbox{\label{fig:tHq_feynman1}}{\includegraphics[width=0.32\textwidth]{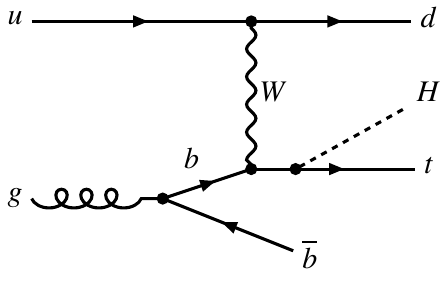}}
\quad
\subcaptionbox{\label{fig:tHq_feynman2}}{\includegraphics[width=0.32\textwidth]{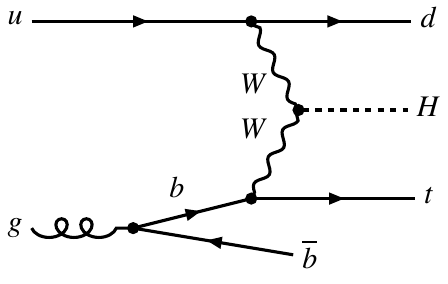}}
\quad
\subcaptionbox{\label{fig:tWH_feynman}}{\includegraphics[width=0.26\textwidth]{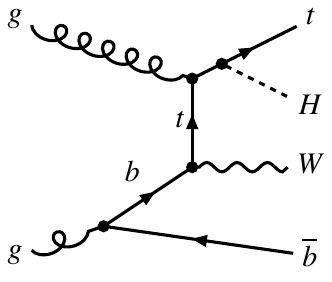}}
\caption{Representative Feynman diagrams for (a) and (b) \tHq and (c) \tWH production at
LO, in the four-flavour scheme.}
\label{fig:feynman_diagrams}
\end{figure}

Example Feynman diagrams of the \tH production process at leading order (LO) are shown in \cref{fig:feynman_diagrams}.
\Cref{fig:tHq_feynman1,fig:tHq_feynman2} show the dominant $t$-channel production mode (\tHq),
while \cref{fig:tWH_feynman} shows the \tWH channel
where the Higgs boson couples to the top quark.
The third electroweak (EW) production mode,
namely the $s$-channel process ($qq^\prime\to W^*\to tHb$),
is expected to have a negligible cross-section at the Large Hadron Collider (LHC) and is neglected.
The graphs shown correspond to the four-flavour scheme (4FS),
where the \Pqb-quark in the initial state comes from gluon splitting
and is not included in the parton distribution function (PDF) of the proton.
The \tHq event topology is characterised by two heavy objects, the
top quark and the Higgs boson, plus an energetic light-flavour quark
(dominantly a \Pqd- or a \Paqu-quark), referred to as the spectator quark, and an extra, mostly soft,
\bquark from gluon splitting. The \tWH event topology is characterised by three heavy objects, the
top quark, the Higgs boson and a \PW boson and an extra, mostly soft, \Pqb-quark
from gluon splitting.
This topology is very similar to that of \ttH production.
In addition, the expected \tWH cross-section is about five times smaller than that of \tHq.
Hence, while the signal is defined as \tHchans,
the analysis is aimed more towards separating \tHq production from the backgrounds.

Under the Standard Model (SM) hypothesis %
the \tH cross-section is strongly suppressed, due to destructive interference between the
diagrams in which the Higgs boson couples to the top quark and those where it couples to the \PW boson~\cite{Demartin:2015uha},
which makes \tH production a sensitive probe of potential new physics beyond the SM,
such as a $CP$-odd mixture in the Higgs--top Yukawa coupling.

The ATLAS Collaboration recently published a search for \tH production in the \Hgg decay mode
with \lumiRunThree of proton--proton collision data
at a centre-of-mass energy of \(\sqrt{s} = \qty{13.6}{\TeV}\)~\cite{HIGP-2024-24}.
The measured signal strength is \(\mutH = \mutHgamgamRunThree\),
with an observed (expected) upper limit at the \qty{95}{\%} confidence level (CL) of
\limtHgamgamRunThreeObs\,(\limtHgamgamRunThreeExp) times the SM prediction.
An earlier search in the same decay channel
using the full \Run{2} data set at \(\sqrt{s} = \qty{13}{\TeV}\)~\cite{HIGG-2020-16}
measured a signal strength of \(\mutH=\mutHgamgam\),
with an observed (expected) upper limit at the \qty{95}{\%} CL
of \limtHgamgamObs\,(\limtHgamgamExp) times the SM prediction.
Using the same full \Run{2} data set, the ATLAS Collaboration also searched for \tH production
in final states either containing two or three charged light leptons (\Pe or \Pgm)
and no hadronically-decaying \Pgt-leptons (\tauhad),
or satisfying selection criteria targeting \Hbb~\cite{HIGP-2024-03}.
After correcting an issue affecting the fake-lepton contribution in that analysis,
the measured signal strength is \(\mutH=\mutHMLbb\),
with an observed (expected) upper limit at the \qty{95}{\%} CL of \limtHMLbbObs\,(\limtHMLbbExp)
times the SM prediction.
The correction has a minor impact on the reported results, affecting only the last quoted digit.

Similar searches using the \Run{2} data set have been published by the CMS Collaboration.
The most recent publications~\cite{CMS-HIG-19-008,CMS-HIG-19-011} target the production of a Higgs boson in association
with either one or two top quarks.
In the multi-lepton final state, the measured signal strength is
\mutHCMSML.
In the \Hbb final state, the observed (expected) upper limit relative to the SM prediction is \num{14.6} (\num{19.3}) at the \qty{95}{\%} CL\@.
In the recent combined measurement of Higgs boson production from the CMS Collaboration~\cite{CMS-HIG-21-018},
\mutH is measured to be \mutHCMScomb.

This publication uses the same \Run{2} data set and complements the previous ATLAS publications~\cite{HIGP-2024-03,HIGG-2020-16}
by selecting events containing either
two charged light leptons and one \tauhad (\dileptau) or
one charged light lepton and two \tauhad (\lepditau) in the final state,
thus targeting semileptonic top quark and \Htautau decays.
These measurements are combined with those in the previous publications of the ATLAS Collaboration
to improve the sensitivity to the \tH production process.

This document is structured as follows: \Cref{sec:atlas} gives a brief overview of the ATLAS detector,
while \cref{sec:data_mc} describes the data and simulated event samples used in the analysis.
The event reconstruction and selection are summarised in \cref{sec:selection}.
This section also includes the description of the neural networks to separate signal from background
and the methods used to estimate the background.
The data-driven techniques used to determine the background due to
reconstructed \tauhad that do not arise from real hadronically decaying \Pgt-leptons
are described in \cref{sec:background}.
The determination of systematic uncertainties is described in \cref{sec:systematics}.
The statistical analysis and the results are presented in \cref{sec:results},
including their combination with previous results from the ATLAS Collaboration.
Finally, conclusions are presented in \cref{sec:conclusion}.


%

%
%

%
\newcommand{\AtlasCoordFootnote}{%
ATLAS uses a right-handed coordinate system with its origin at the nominal interaction point (IP)
in the centre of the detector and the \(z\)-axis along the beam pipe.
The \(x\)-axis points from the IP to the centre of the LHC ring,
and the \(y\)-axis points upwards.
Polar coordinates \((r,\phi)\) are used in the transverse plane,
\(\phi\) being the azimuthal angle around the \(z\)-axis.
The pseudorapidity is defined in terms of the polar angle \(\theta\) as \(\eta = -\ln \tan(\theta/2)\) and is equal to the rapidity
$ y = \frac{1}{2} \ln \left( \frac{E + p_z}{E - p_z} \right) $ in the relativistic limit.
Angular distance is measured in units of \(\Delta R \equiv \sqrt{(\Delta y)^{2} + (\Delta\phi)^{2}}\).}

\section{ATLAS detector}%
\label{sec:atlas}

The ATLAS experiment~\cite{PERF-2007-01} at the LHC is a multipurpose particle detector
with a forward--backward symmetric cylindrical geometry and a near \(4\pi\) coverage in
solid angle.\footnote{\AtlasCoordFootnote}
It consists of an inner tracking detector surrounded by a thin superconducting solenoid
providing a \qty{2}{\tesla} axial magnetic field, electromagnetic and hadronic calorimeters, and a muon spectrometer.
The inner tracking detector covers the pseudorapidity range \(|\eta| < 2.5\).
It consists of silicon pixel, silicon microstrip, and transition radiation tracking detectors.
Lead/liquid-argon (LAr) sampling calorimeters provide electromagnetic (EM) energy measurements
with high granularity within the region \(|\eta|< 3.2\).
A steel/scintillator-tile hadronic calorimeter covers the central pseudorapidity range (\(|\eta| < 1.7\)).
The endcap and forward regions are instrumented with LAr calorimeters
for EM and hadronic energy measurements up to \(|\eta| = 4.9\).
The muon spectrometer surrounds the calorimeters and is based on
three large superconducting air-core toroidal magnets with eight coils each.
The field integral of the toroids ranges between \num{2.0} and \qty{6.0}{\tesla\metre}
across most of the detector.
The muon spectrometer includes a system of precision tracking chambers up to \(|\eta| = 2.7\) and fast detectors for triggering up to \(|\eta| = 2.4\).
The luminosity is measured mainly by the LUCID--2~\cite{LUCID2} detector which is located close to the beampipe.
A two-level trigger system was used to select events~\cite{TRIG-2016-01}.
The first-level trigger is implemented in hardware and used a subset of the detector information
to accept events at a rate close to \qty{100}{\kHz}.
This is followed by a software-based trigger that
reduced the accepted rate of complete events to \qty{1.25}{\kHz} on average
depending on the data-taking conditions.
A software suite~\cite{SOFT-2022-02} is used in data simulation, in the reconstruction
and analysis of real and simulated data, in detector operations, and in the trigger and data acquisition
systems of the experiment.


%

%
%

%

%
\section{Data and simulated event samples}%
\label{sec:data_mc}

This analysis uses data from \pp collisions at $\sqrt{s} = \qty{13}{\TeV}$ collected with the ATLAS detector
at the LHC during 2015--2018.
Stringent detector and data-quality requirements are applied~\cite{DAPR-2018-01},
resulting in a data sample corresponding to a total integrated luminosity of \lumi.
The events were selected by single-lepton triggers with variable electron and muon transverse momentum (\pT)
thresholds, and with various identification and isolation criteria
depending on the lepton flavour and the data-taking period~\cite{TRIG-2018-05, TRIG-2018-01}.
The lowest $\pT$ threshold at trigger level used for electrons was \qty{24}{\GeV} (\qty{26}{\GeV}),
while for muons the threshold was \qty{20}{\GeV} (\qty{26}{\GeV}) in 2015 (2016--2018).

The simulated event samples are identical to those used in the ATLAS \Run{2} \tH search of Ref.~\cite{HIGP-2024-03}
and more details of their generation can be found there.
They are summarised in \Cref{tab:mc-samples}.
Depending on the process, the simulation of the hard-scatter process was performed at
LO or next-to-leading order (NLO) in perturbative QCD\@.
The PDF sets used were chosen to match the order of the matrix element (ME) calculations
and can be at LO, NLO or next-to-next-to-leading order (NNLO).
To simulate the effects of additional \(\Pp{}\Pp\) collisions in the same, next or previous bunch crossings (\pileup),
additional interactions were generated using the simulated minimum-bias events provided by \PYTHIA[8.186]~\cite{Sjostrand:2007gs}
with the A3 tune~\cite{ATL-PHYS-PUB-2016-017} and the MSTW2008LO PDF set~\cite{Martin:2009iq}, and were overlaid onto each simulated hard-scatter event.
The Monte Carlo (MC) samples were reweighted so that the pile-up distribution matches the one observed in the data.

\begin{table}[htbp]
\tablesetup
\setlength\tabcolsep{1.5pt}
\caption{Summary of the configurations used to generate events for signal (\tHq\ and \tWH) and background processes.
\enquote{ME order} refers to the order and specific corrections for the matrix element calculation,
\enquote{Tune} to the set of parameters used in the parton shower model
and \enquote{\(V\)} to a vector boson (\PW\ or \PZ).
}
\label{tab:mc-samples}
\begin{tabular}{ l l l l l l }
\toprule
Process & Generator & ME order & Parton shower & PDF set & Tune \\
\midrule
\tHq                    & MG5\_aMC~\cite{Alwall:2014hca}  & NLO & \PYTHIA[8]~\cite{Sjostrand:2007gs, Sjostrand:2014zea} & \NNPDF[3.0nlo]~\cite{Ball:2014uwa} & A14~\cite{ATL-PHYS-PUB-2014-021} \\
\tWH                    & MG5\_aMC & NLO & \PYTHIA[8]\ & \NNPDF[3.0nlo] & A14 \\
\midrule
\ttbar                  & \POWHEGBOX~\cite{Frixione:2007nw,Nason:2004rx,Frixione:2007vw,Alioli:2010xd} & NLO & \PYTHIA[8] & \NNPDF[3.0nlo] & A14  \\
\ttH                    & \POWHEGBOX  & NLO & \PYTHIA[8]\ & \NNPDF[3.0nlo] & A14 \\
\ttW                    & \Sherpa[2.2.10]~\cite{Bothmann:2019yzt} & NLO QCD+EW & \Sherpa &  \NNPDF[3.0nnlo] & default \\
\ttW (EW)               & \Sherpa[2.2.10] & LO & \Sherpa &  \NNPDF[3.0nnlo] & default \\
\ttZ                    & MG5\_aMC & NLO & \PYTHIA[8] & \NNPDF[3.0nlo] & A14  \\
Single top quark        & \POWHEGBOX & NLO & \PYTHIA[8] & \NNPDF[3.0nlo] & A14   \\
($t$-ch., $\Wt$, $s$-ch.) &  &  &  &  \\
$V+\text{jets}$         & \Sherpa[2.2.1] & \MEPSatNLO~\cite{Hoeche:2012yf} & \Sherpa & \NNPDF[3.0nlo] & default \\
$VV$, $qqVV$            & \Sherpa[2.2.2]  & \MEPSatNLO & \Sherpa & \NNPDF[3.0nnlo] & default  \\
$VVV$                   & \Sherpa[2.2.2]  & \MEPSatNLO & \Sherpa & \NNPDF[3.0nnlo] & default  \\
\tZq                    & MG5\_aMC & NLO & \PYTHIA[8] & \NNPDF[3.1nlo]~\cite{NNPDF:2017mvq} & A14 \\
\tWZ                    & MG5\_aMC & NLO & \PYTHIA[8] & \NNPDF[2.3lo]~\cite{Ball:2012cx} & A14 \\
$t\bar t t, t\bar t t\bar t$      & MG5\_aMC & QCD NLO & \PYTHIA[8] & \NNPDF[3.1nlo] & A14 \\
$\ggH, \qqH$ & \POWHEGBOX & QCD NLO & \PYTHIA[8] & \CTEQ[6L1]~\cite{Pumplin:2002vw} & \AZNLO~\cite{STDM-2012-23} \\
$\WH, \ZH$ & \PYTHIA[8] & QCD NLO & \PYTHIA[8] & \NNPDF[2.3lo] & A14 \\
\bottomrule
\end{tabular}
\end{table}


The simulated \tHq signal samples were generated in the 4FS
and are normalised to the NLO QCD theory prediction without NLO EW corrections~\cite{Maltoni_2012}.
At \(\rts = \qty{13}{\TeV}\) and in the five-flavour scheme (5FS),
where the \Pqb-quark is treated as a massless parton in the proton sea,
the predicted cross-section is
\(74.3^{+6.5\%}_{-14.9\%}~(\text{QCD scale})~\pm~3.7\%~(\text{PDF+\alphas})\,\unit{\fb}\)~\cite{deFlorian:2016spz,YR4twiki}.
The simulated \tWH\ signal samples were also generated in the 4FS and are normalised to the NLO QCD theory prediction in
the 5FS~\cite{deFlorian:2016spz,YR4twiki} and used the
DR2 scheme~\cite{Demartin:2016axk} to treat the overlap between
the \tWH\ and \ttH\ processes.
The predicted cross-section is $\sigma(\tWH)_{\text{NLO}} = 15.2^{+4.9\%}_{-6.7\%}~(\text{QCD scale})~\pm~6.3\%~(\text{PDF+\alphas})\,\unit{\fb}$~\cite{deFlorian:2016spz,YR4twiki}.

Most of the simulated samples used for the nominal signal and background estimates
were processed through a simulation~\cite{SOFT-2010-01}
of the ATLAS detector based on \GEANT~\cite{Agostinelli:2002hh},
while those used for evaluating systematic uncertainties were processed with a fast simulation~\cite{SOFT-2010-01}
that relies on a parameterisation of the calorimeter response~\cite{ATL-PHYS-PUB-2010-013}.
In order to have enough events for the multivariate methods used,
fast-simulated event samples are also used for certain nominal samples,
particularly for the \tHq and \tWH signal processes and the \tttt background process, with no loss of accuracy.


%

%
%

%

%
\section{Event reconstruction and selection}%
\label{sec:selection}

The expected signal-to-background ratio is very small.
Hence a multivariate analysis using artificial neural networks (NN) is used to enhance the sensitivity.
The criteria used to select the events for the NN training (denoted \enquote{pre-selection}) are described first,
followed by a description of the architecture and training of the NNs used to separate the \tH signal from the backgrounds.
The training uses \tHq and \tWH events as signal and all other processes as background.

The reconstruction of leptons\footnote{%
In the following, \enquote{leptons} refers to all charged leptons (\Pe, \Pgm and \Pgt),
while \enquote{light leptons} means electrons or muons.}
and jets used in this document is identical
to that used in Ref.~\cite{HIGP-2024-03} and is therefore only summarised here.
\Tight electrons have to fulfil the \texttt{tight} lepton identification criteria~\cite{EGAM-2018-01}
and the \texttt{tight} isolation working point~\cite{HIGG-2017-02}.
They are required to satisfy \(\pt > \qty{10}{\GeV}\) and \(\abs{\eta } <2.47\),
excluding the transition region between the endcap and barrel calorimeters ($1.37<|\eta|<1.52$).
\Tight muons have to fulfil the \texttt{medium} muon identification criteria~\cite{MUON-2018-03,MUON-2022-01}
and the \texttt{tight} isolation working point~\cite{HIGG-2017-02}.
They are required to satisfy $\pt > \SI{10}{\GeV}$ and $|\eta| < 2.5$.
\Loose light leptons have to satisfy the same selection criteria,
except that the isolation criteria are not applied.
\Loose electrons also do not have to satisfy
the electron charge-misassignment veto and the electron conversion candidate veto~\cite{EGAM-2018-01}.
\Tight \tauhad are required to have one or three associated tracks, electric charge of $\pm 1$,
$\pT > \qty{20}{\GeV}$, and $\abs{\eta} < 2.5$, excluding the barrel--endcap transition region (\(1.37 < \abs{\eta} < 1.52\)).
Each such \tauhad must satisfy the \texttt{medium} RNN identification selection requirement,
with efficiencies of \qty{75}{\%} for 1-prong\footnote{%
A \tauhad is classified according to the number of its charged decay particles, here referred to as prongs.}
and \qty{60}{\%} for 3-prong \tauhad.
One-prong \tauhad have to also satisfy the \texttt{loose} \textsc{eBDT} requirement,
with an efficiency of 95\%~\cite{ATLAS-CONF-2017-029, ATL-PHYS-PUB-2022-044}.
\Loose \tauhad have to satisfy the same criteria, but with the \texttt{loose} RNN identification requirement.

Each event is required to have $\geq 2$ jets (maximum 6) with \(\pT > \qty{20}{\GeV}\) and \(\abs{\eta} \leq 4.5\).
The jets are reconstructed using the \antikt algorithm with a radius parameter of \(R=0.4\)~\cite{Cacciari:2008gp,Fastjet}.
To reduce the effect of \pileup, %
each jet with \(\pT < \qty{60}{\GeV}\) and \(\abs{\eta} < 2.4\) must satisfy
the \texttt{tight} working point of the jet vertex tagger (JVT)~\cite{PERF-2014-03}.
Jets within \(2.5 < \abseta < 4.5\) and \(\pT < \qty{120}{\GeV}\)
must satisfy the \texttt{tight} working point of the forward jet vertex tagger (fJVT)~\cite{PERF-2016-06} criteria.
In addition, one or two of these jets must be identified as a \bjet
using the \qty{70}{\%} working point of the \texttt{DL1r} algorithm~\cite{FTAG-2019-07}.
An overlap removal procedure is applied to avoid double-counting of objects,
following the one described in Ref.~\cite{HIGP-2024-03}.

The missing transverse momentum \MpT (with magnitude \MET)
is defined as the negative vector sum of the transverse momenta
of all identified and calibrated objects and remaining unclustered energy,
the latter of which is estimated from low-\pT tracks associated with the primary vertex but not assigned to any lepton or jet candidate~\cite{PERF-2016-07}.
The \MET is taken as a measure of the undetectable particles, and is affected by energy losses due to detector inefficiencies and acceptance, and by energy resolution.

\subsection{\dileptausec channel}%
\label{sec:selection:dileptau}

The pre-selection in the \dileptau channel requires one \tight \tauhad and two \tight light leptons,
where the leading light lepton has to be matched to a trigger object and have \(\pT > \qty{27}{\GeV}\).
The sum of the electric charges of the leptons must be \(\pm 1\).
The events are separated into same-sign (\SS) and opposite-sign (\OS) categories based on the electric charges of the two light leptons.
The \SS category is used in the NN training and as the signal region, as it provides the best signal-to-background ratio,
while the \OS category is used to define control regions for the estimation of backgrounds with fake \tauhad.
Fake \tauhad refers to objects such as jets that are falsely identified as a \tauhad.

The dominant background processes in the \dileptau channel are reducible and mostly contain a fake \tauhad,
including \ttbar and \Zjets production.
Further sources of backgrounds include \diboson\ and \ttX processes (\ttH, \ttZ and \ttW).
For events with a \SS pair all background contributions are reduced significantly,
and the dominant backgrounds arise from \ttbar and \ttX production.
The expected event yields after the pre-selection in the \dilepSStau and \dilepOStau channels are given in \cref{tab:yields:prefit}.

To reconstruct the \tHq topology, the leptons have to be associated with the Higgs boson or the top quark.
In most \tHq signal events the Higgs boson will have decayed into two \Ptau-leptons of which one decays hadronically.
Since \qty{85}{\%} of the \tauhad in the signal events originate from the Higgs boson decay,
it is assumed that one light lepton originates from the Higgs boson and the other from the top quark.
Events in which the \tauhad originates from the top quark are still counted as signal.
In the \SS events, a boosted decision tree (BDT) is used to assign the leading light lepton to either the top quark or the Higgs boson;
the sub-leading light lepton is then assigned to the other particle.
The discriminating variables used in the training
comprise angular distances and masses relating the \tauhad and the light leptons,
as well as similar quantities between the light leptons and the leading \pT \btagged jet.
The training is based on a five-fold cross-validation, and provides stable results with no visible overtraining.
A selection requirement on the BDT output is used to decide which of the two possible assignments to take.
As a result, approximately \qty{88}{\%} of the leptons are correctly assigned,
while the remaining \qty{12}{\%} are incorrectly assigned.
For the \OS events, the light lepton with charge opposite that of the \tauhad is assumed to originate from the Higgs boson.

\begin{table}[htbp]
\tablesetup
\sisetup{table-align-uncertainty}%
\caption{Expected event yields in the \dilepSStau, \dilepOStau and \lepditau channels,
using the pre-selection criteria.
For the \dileptau channels, the single top \tchan contribution is included in \enquote{Minor bkgs.}, as it is very small.
The uncertainties correspond to the statistical uncertainties of the simulated samples.
}
\label{tab:yields:prefit}
\begin{tabular}{l |
S[table-format = 5.2(4), separate-uncertainty]
S[table-format = 5.2(4), separate-uncertainty]
S[table-format = 4.2(4), separate-uncertainty]
}
\toprule
Process & \multicolumn{3}{c}{Expected event yields} \\
& \multicolumn{1}{c}{\dilepSStau}
& \multicolumn{1}{c}{\dilepOStau}
& \multicolumn{1}{c}{\lepditau} \\
\midrule
\tHq            & 1.08 \pm 0.03  & 1.67 \pm 0.04   & 2.11 \pm 0.04  \\
\tWH            & 0.53 \pm 0.03  & 1.37 \pm 0.05   & 0.92 \pm 0.04  \\
\tWZ            & 1.99 \pm 0.03  & 15.67 \pm 0.07  & 3.23 \pm 0.03  \\
\ttbar          & 23.5 \pm 1.8   & 5291 \pm 27     & 805 \pm 11     \\
\ttW            & 30.44 \pm 0.43 & 67.63 \pm 0.64  & 9.99 \pm 0.24  \\
\ttZ            & 15.00 \pm 0.33 & 113.00 \pm 0.94 & 22.29 \pm 0.48 \\
\ttH            & 17.46 \pm 0.14 & 44.64 \pm 0.19  & 25.8 \pm 0.19  \\
\tZq            & 4.50 \pm 0.12  & 33.67 \pm 0.33  & 9.97 \pm 0.18  \\
Single top (\tW) & 1.27 \pm 0.40 & 246.1 \pm 5.6   & 37.1 \pm 2.2   \\
Single top (\(t\)-ch.) & {--}    & {--}            & 4.60 \pm 0.50  \\
\Zjets          & 1.05 \pm 0.21  & 4666 \pm 37     & 56.9 \pm 4.4   \\
\Wjets          & 1.5 \pm 1.1    & 0.70 \pm 0.32   & 47.6 \pm 6.1   \\
\Diboson        & 8.51 \pm 0.33  & 162.8 \pm 2.3   & 19.98 \pm 0.74 \\
Minor bkgs.     & 1.14 \pm 0.30  & 12.5 \pm 2.3    & 0.65 \pm 0.10  \\
\midrule
Total           & 108.0 \pm 2.2  & 10657 \pm 46    & 1046 \pm 13    \\
\bottomrule
\end{tabular}

\end{table}

\subsection{\lepditausec channel}%
\label{sec:selection:lepditau}

In the \lepditau channel the pre-selection requires two \tight \tauhad and one \tight light lepton,
where the light lepton has to be matched to a trigger object and have \(\pT > \qty{27}{\GeV}\).
No electric charge requirement is imposed on the light lepton,
while the two \tauhad must have opposite electric charge.
In addition, a selection on \(\Delta\phi\) between the two \tauhad is applied
(\(\Delta\phi_{\tauhad,\tauhad}<3.05\)) to all regions
to lower the mismodelled \Wjets fraction.
In \qty{86}{\%} of the \tHq signal events the assumption that the light lepton originates from the top quark is correct.

The \lepditau channel is dominated by reducible backgrounds due to one or two fake \tauhad,
including \ttbar/\tW, \Zjets and \Wjets events.
Important irreducible backgrounds are \ttZ, \ttH, \diboson and \tZq processes.
An overview of the expected event yields is given in \cref{tab:yields:prefit}.

\subsection{Top-quark reconstruction}%
\label{sec:selection:top}

It is advantageous to reconstruct the kinematics of both the top quark and the Higgs boson
to suppress background.
However, an accurate reconstruction of these particles is challenging,
due to the presence of typically three (\lepditau) or four (\dileptau) neutrinos in the final state,\footnote{%
This counting ignores neutrinos in jets containing a \Pqb-hadron (\bjets).}
of which two or three are from the Higgs boson and one is from the top quark. %
The strategy used to fully reconstruct the final state is
to reconstruct the top quark first and then the $\Pgt\Pgt$ system.
This idea is based on the fact that total missing transverse momentum, \MpT,
is the sum of the missing transverse momentum from the $\Pgt\Pgt$ system and that from the top quark.
As the total \MpT is measured through the standard object reconstruction procedure~\cite{PERF-2016-07},
if the missing transverse momentum due to the top-quark decay products (\MpT(top)) is known,
\MpT(Higgs) can be calculated using
\(\MpT(\text{Higgs}) = \MpT - \MpT(\text{top})\).
This strategy has the drawback that \MpT(Higgs) depends on \MpT(top),
which means that the accuracy with which the neutrino from the top quark is reconstructed
affects the accuracy of the reconstructed invariant mass of the Higgs boson.

The leading \pT \Pqb-quark jet, assumed to originate from the top-quark decay,
and the charged light lepton are reconstructed in the detector.
Studies show that, to a reasonable approximation, the \pT of the neutrino, \(\pTX[\Pgn, \text{top}]\),
is inversely proportional to the \pT of the charged light lepton, \(\pTX[\Pl, \text{top}]\),
\begin{math}
\pTX[\Pgn, \text{top}] \approx \qty{1600}{\GeV^{2}} / \pTX[\Pl, \text{top}]\,.
\end{math}
It is also found that the azimuthal angle $\phi$ of the neutrino
tends to be perpendicular to that of the charged light lepton.
Using this information \MpT(top) can be calculated and used to estimate the \MpT(Higgs).

\subsection{Higgs boson mass reconstruction}%
\label{sec:selection:higgs}

The missing mass calculator (MMC) method~\cite{ELAGIN2011481}
is used to estimate the reconstructed Higgs boson mass.
This method uses the measured momenta of the visible \Pgt decay products
and the \MpT(Higgs) to reconstruct the mass of the \Pgt{}\Pgt system.
The algorithm performs a likelihood scan over the unknown kinematic parameters of the \Pgt decays,
constrained by the measured \MpT(Higgs).
The mass of the \Pgt{}\Pgt system is then taken from the hypothesis with the highest likelihood value.
The resulting peak of the mass distribution is close to the measured value,
while the width of the peak is similar to that obtained for the visible mass,
which uses the directly measured decay products only.

\subsection{Neural network training}%
\label{sec:NN-training}

\begin{figure}[htbp]
\centering
\subcaptionbox{}{%
\includegraphics[width=0.57\textwidth]{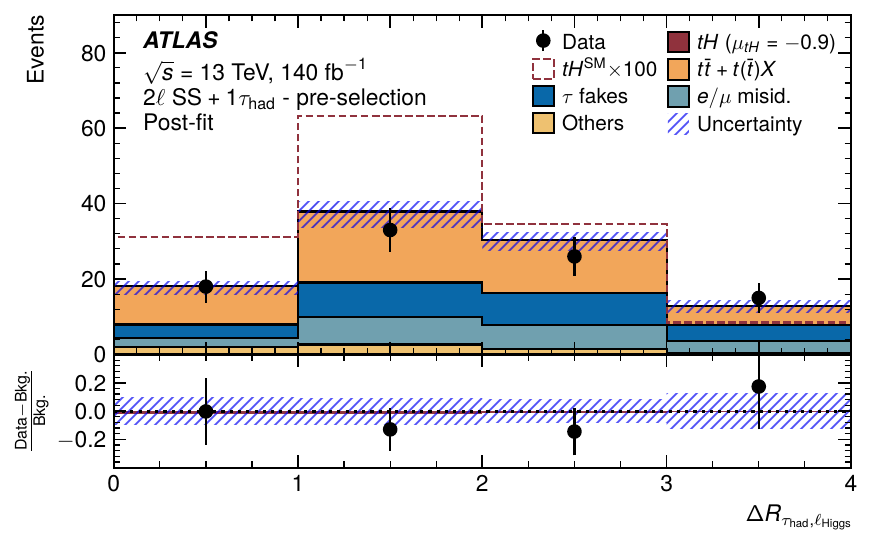}
}
\quad
\subcaptionbox{\label{fig:dilepSStau:NN:features:tau_pt}}{%
\includegraphics[width=0.57\textwidth]{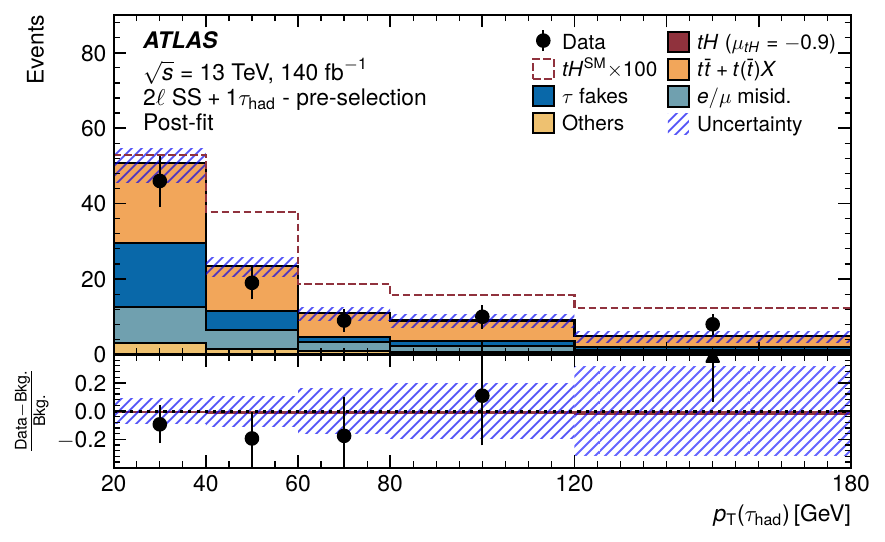}
}
\quad
\subcaptionbox{}{%
\includegraphics[width=0.57\textwidth]{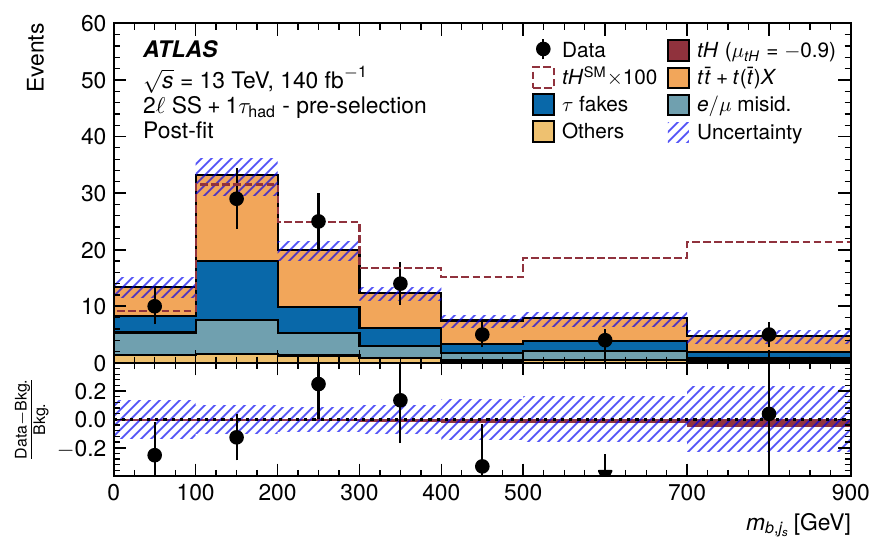}
}
\caption{Example features used in the NN training in the \dilepSStau channel:
(a) \(\DeltaR_{\tauhad, \Pl_{\text{Higgs}}}\) is the angular distance between the \tauhad and the light lepton assigned to the Higgs boson,
(b) \(\pT(\tauhad)\) is the transverse momentum of the \tauhad,
and (c) \(m_{\Pqb, \js}\) is the invariant mass of the \bjet associated with the top quark and the spectator jet.
\(\Pqt(\Paqt)X\) represents all processes, except the signal,
with one or two top quarks and a heavy boson (\PW, \PZ, \PH) in the final state.
Overflows are included in the last bin of each distribution.
The event fractions correspond to those determined in the fit discussed in \cref{sec:results}.
The uncertainties shown are the total uncertainties in the data and MC simulation.
}%
\label{fig:dilepSStau:NN:features}
\end{figure}

\begin{figure}[htbp]
\centering
\subcaptionbox{}{%
\includegraphics[width=0.57\textwidth]{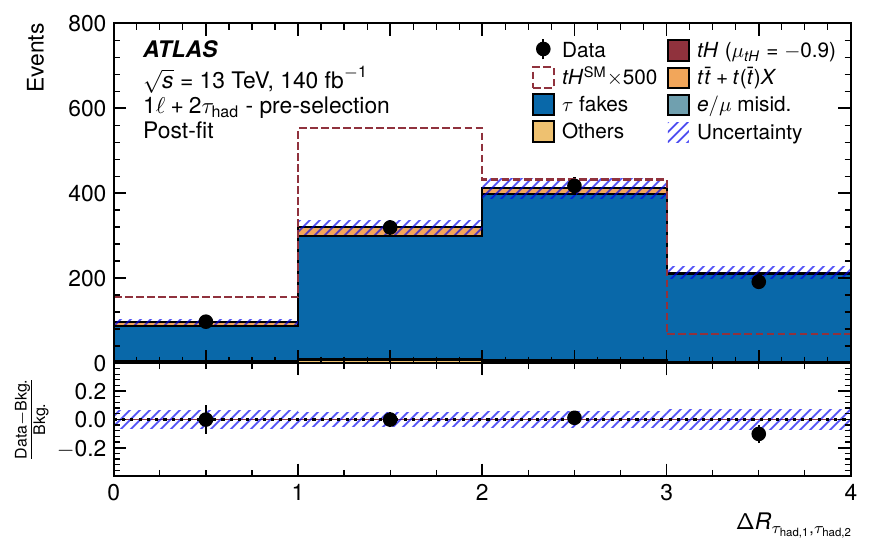}
}
\quad
\subcaptionbox{\label{fig:lepditau:NN:features:tau_pt}}{%
\includegraphics[width=0.57\textwidth]{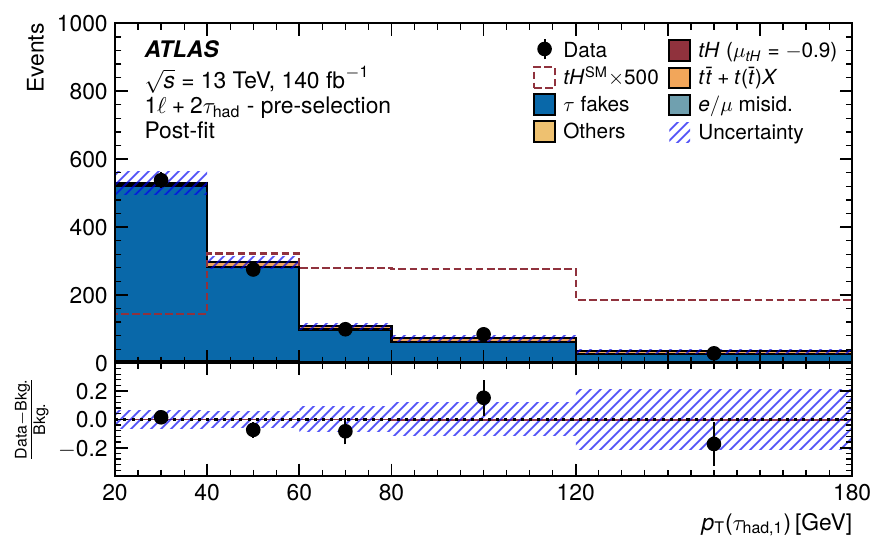}
}
\quad
\subcaptionbox{}{%
\includegraphics[width=0.57\textwidth]{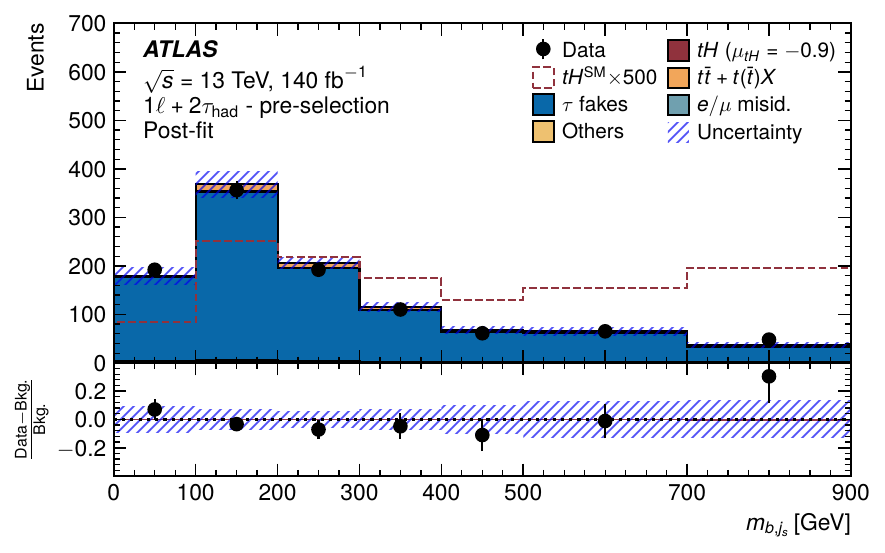}
}
\caption{Example features used in the NN training in the \lepditau channel:
(a) \(\DeltaR_{\Pgt_{\text{had},1}, \Pgt_{\text{had},2}}\) is the angular distance between the two \tauhad,
(b) \(\pT(\Pgt_{\text{had,1}})\) is the transverse momentum of the leading \tauhad,
and (c) \(m_{\Pqb, \js}\) is the invariant mass of the \bjet associated with the top quark and the spectator jet.
\(\Pqt(\Paqt)X\) represents all processes, except the signal,
with one or two top quarks and a heavy boson (\PW, \PZ, \PH) in the final state.
Overflows are included in the last bin of each distribution.
The event fractions correspond to those determined in the fit discussed in \cref{sec:results}.
The uncertainties shown are the total uncertainties in the data and MC simulation.
}%
\label{fig:lepditau:NN:features}
\end{figure}

Tensorflow Keras-based neural networks~\cite{Keras} are used for separation of the signal from the background processes.
Separate networks are trained in the \dilepSStau and \lepditau regions,
using the simulated events that satisfy the pre-selection criteria defined in \cref{sec:selection}.
Kinematic variables of the final state products and the \MET are used as features.
A feature of the \tHq process is the spectator jet, \js,
which originates from the underlying \(t\)-channel production.
The spectator jet is defined as the
non-\btagged jet that has the highest invariant mass when combined with the \btagged jet associated with the top quark.
In particular, the $\abseta$ of the spectator jet is typically higher than that of other jets.
Examples of input variables used in the training are shown in
\cref{fig:dilepSStau:NN:features} for the \dilepSStau channel
and in \cref{fig:lepditau:NN:features} for the \lepditau channel.
The complete list of features is given in \cref{sec:NN-features}.

Hyperparameter optimisation is done using an evolutionary process followed by a fine grid search.
The best hyperparameters in each channel are selected based on a five-fold cross-validation,
using three out of the five folds to train a model.
The optimal hyperparameters are subsequently selected based on this model's predictions on the events of the fourth fold.
The fifth fold is not used in model building nor in evaluation.
Predictions on this fifth fold are used solely in the profile likelihood fit region definition.
This ensures that the predictions of the simulated events in the fit are only executed by models
that have never seen those particular events,
neither in training nor in hyperparameter optimisation.

During the training, the events are split into three classes,
which are reweighted to each contain the same number of events.
For the \dilepSStau channel these are: \tH signal; \ttbar and \(\ttbar X\),
where \(X\) is \PW, \PZ or \PH; other backgrounds.
In the \lepditau channel the categories are: \tH signal; \ttbar; other backgrounds.
One categorical classifier is trained in each channel,
with three output nodes, corresponding to the three categories.
The output score distributions of the NNs trained with \tH as signal, \ONNtH,
are shown in \cref{fig:SR_SM} for both the \dilepSStau and \lepditau channels,
after performing the fit described in \cref{sec:results}.

%
%
%
%
%
%
%
%
%


%

%
%

%

%
\section{Background estimation and fit region definition}
\label{sec:background}

Accurately simulating the probability to misidentify a light lepton\footnote{%
Misidentified light leptons are those that do not originate from the decay of a \PW or \PZ boson, i.e.\ are not prompt leptons,
but rather from other sources such as heavy-flavour hadron decays.}
or fake a \tauhad is challenging, so a data-driven approach is used.
The overall rates of backgrounds due to misidentified and fake leptons are estimated
using templates from simulated events in the final profile likelihood fit,
which is described in \cref{sec:results}.
This takes advantage of the high statistics available in the simulated samples.
The method involves constructing templates for each object type (electron, muon, or \tauhad)
and determining the normalisation of the background templates by matching their rates to the data
in designated background-enriched control regions (CRs).
Several CRs are defined to provide regions dominated by different processes.

In the following sections, the normalisation factors (NF) used in the estimates
of the backgrounds due to misidentified light leptons and fake \tauhad are introduced.
If an event contains more than one such misidentified light lepton or \tauhad fake,
the total normalisation corresponds to the product of the relevant NFs.
At the end of the section, the definitions of the signal (SR) and control regions used in the final profile likelihood fit are given.

\subsection{\dileptausec channel}%
\label{sec:fake:dileptau}

The probability for a \tauhad to originate from a jet depends mainly on the \tauhad measured properties
($\abs{\eta}$, \pT and the number of charged-particle tracks associated with the \tauhad).
While the \(\abs{\eta}\) distribution of the \tauhad shows good agreement between the simulation and data,
a significant mismodelling is observed in \pT.
Thus, to correct the MC simulation for mismodelling of the \tauhad fakes, a \pT binning is used
and separate NFs are determined in each bin.
The choice of \pT binning is driven by the size of data/MC mismodelling,
the available MC simulation statistics,
as well as by the relative contribution of a specific \pT bin in the signal region.
The probability of a \tauhad to originate from a jet
and the dependence of this probability on \pT are observed in simulation
to be very similar for \tauhad reconstructed with one charged-particle track and
\tauhad reconstructed with three charged-particle tracks;
hence the same NFs and \pT binning are used for both types of \tauhad.
The background prediction is found to have little dependence on the choice of \pT-binning;
a parameterisation of the \pT-dependence also produced a similar estimate.

Backgrounds from misidentified light leptons are dominated by heavy-flavour (HF) hadron decays
and are hence labelled as \(\Pe/\Pgm_{\text{HF}}\).
In total five different NFs are used to correct the MC simulation:

\begin{itemize}
\item \ke: applied to all non-prompt electrons in the MC samples, except electrons from photon conversion;
\item \kmu: applied to all non-prompt muons in the MC samples;
\item \ktau{1}: applied to all \tauhad fakes with \(\qty{20}{\GeV} < \pT \leq \qty{30}{\GeV}\);
\item \ktau{2}: applied to all \tauhad fakes with \(\qty{30}{\GeV} < \pT \leq \qty{60}{\GeV}\);
\item \ktau{3}: applied to all \tauhad fakes with \(\pT > \qty{60}{\GeV}\).
\end{itemize}

Leptons with a misidentified charge do not play a significant role in this channel.
However, electrons from \(\gamma\)-conversions are expected to contribute at the \qty{3}{\%} level to the SR\@.
The creation of a sufficiently well populated control region for these electrons is not possible.
Following the estimates made in Ref.~\cite{HIGP-2024-03},
an uncertainty of \qty{30}{\percent} is applied to the predicted rate of events including a \(\gamma\)-conversion.

\subsection{\lepditausec channel}
\label{sec:fake:lepditau}

The dominant background in the \lepditau channel is due to jets being misidentified as \tauhad.
The predicted fraction of reducible backgrounds due to misidentified light leptons is negligible.
Most of the \ttbar background satisfies the analysis selection through jets that fake the sub-leading in \pT \tauhad,
although significant contributions arise from fakes of the highest \pT \tauhad and fakes of both \tauhad.
Other significant backgrounds that pass the selection via a jet faking a \tauhad include \Wjets, \Zjets, and single top (\tW) processes.

A \pT dependence of the ratio of data to MC simulation, caused by \tauhad fakes, is also observed in this channel.
This dependence is consistent between the leading and sub-leading \tauhad.
Therefore, the \pT spectrum is parameterised using three NFs with the same binning as in the \dileptau channel:
\ktau{1}, \ktau{2} and \ktau{3}.

\subsection{Signal and control regions}%
\label{sec:regions}

The normalisations of the different sources of backgrounds due to misidentified or fake leptons
are determined in dedicated control regions.
As discussed above, a \pT dependence of the scale factor needed to normalise
the predicted number of events with \tauhad fakes is observed. Hence, three independent CRs are defined
to correct the MC simulation prediction of the \tauhad fake background in both channels.

The following regions are defined in the \dileptau channel,
and all except the validation region (VR) are used in the profile likelihood fit.
The VR serves to check the background modelling in a region close to the \SR.

\begin{itemize}
\item \SR: two \SS \tight light leptons and one \tight \tauhad;
\item \VR: two \SS light leptons,
where at least one is \loose but not \tight, and a \loose but not \tight \tauhad;
\item \CRe: two \SS light leptons of which at least one is \loose but not \tight and one \tight \tauhad.
The sub-leading light lepton has to be an electron;
\item \CRmu: two \SS light leptons of which at least one is \loose but not \tight and one \tight \tauhad.
The sub-leading light lepton has to be a muon;
\item \CRtau{1}: two \OS \tight light leptons and one \tight \tauhad with \(\qty{20}{\GeV} < \pT \leq \qty{30}{\GeV}\),
\(\mll < \qty{66}{\GeV}\) or \(\mll > \qty{116}{\GeV}\) where \(\mll\) is the invariant mass of the two light leptons;
\item \CRtau{2}: two \OS \tight light leptons and one \tight \tauhad with \(\qty{30}{\GeV} < \pT \leq \qty{60}{\GeV}\),
\(\mll < \qty{66}{\GeV}\) or \(\mll > \qty{116}{\GeV}\);
\item \CRtau{3}: two \OS \tight light leptons and one \tight \tauhad with \( \pT > \qty{60}{\GeV}\),
\(\mll < \qty{66}{\GeV}\) or \(\mll > \qty{116}{\GeV}\).
\end{itemize}

\begin{figure}[htbp]
\centering
\subcaptionbox{}{%
\includegraphics[width=0.45\textwidth]{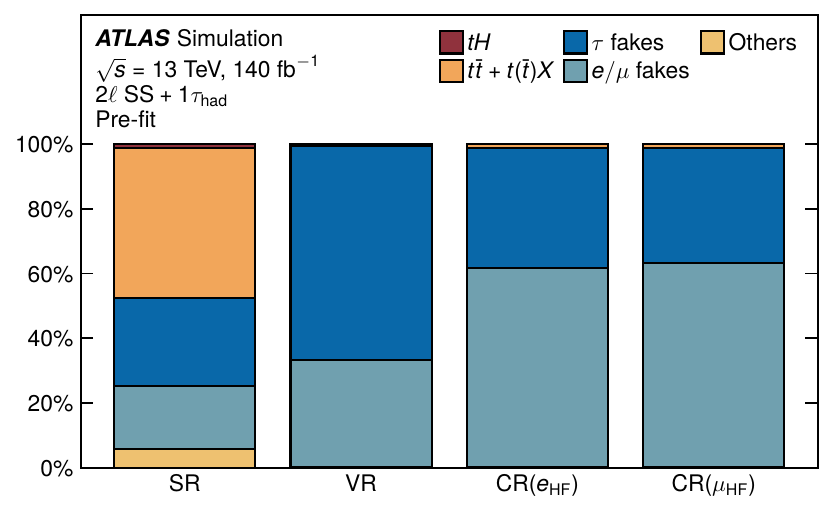}
}
\quad
\subcaptionbox{}{%
\includegraphics[width=0.45\textwidth]{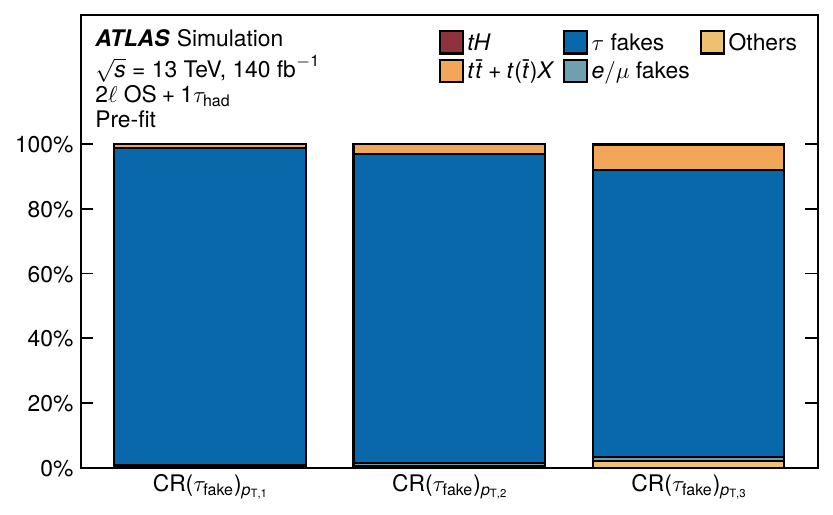}
}
\caption{Composition of the SR, the VR and the CRs used as input to the \dileptau profile likelihood fit:
(a) in the \dilepSStau channel and (b) in the \dilepOStau channel.
\(\Pqt(\Paqt)X\) represents all processes, except the signal,
with one or two top quarks and a heavy boson (\PW, \PZ, \PH) in the final state.
}
\label{fig:fake_yields:dileptau}
\end{figure}

The selection on \mll\ is applied to the \tauhad fake CRs to lower the \Zjets fraction in these regions,
as this selection is also applied in the SR\@.
The composition of the different regions used in the profile likelihood fit is summarised in \cref{fig:fake_yields:dileptau}.

Similar regions are defined in the \lepditau channel:

\begin{itemize}
\item \SR: one \tight light lepton and two \tight \tauhad and \(\ONNtH > 0.45\);
\item \VR: one \tight light lepton and two \tight \tauhad and \(\ONNtH < 0.45\);
\item \CRtau{1}: one \tight light lepton, a \loose but not \tight \tauhad with \(\qty{20}{\GeV} < \pT \leq \qty{30}{\GeV}\) and a \tight \tauhad;
\item \CRtau{2}: one \tight light lepton, a \loose but not \tight \tauhad with \(\qty{30}{\GeV} < \pT \leq \qty{60}{\GeV}\) and a \tight \tauhad;
\item \CRtau{3}: one \tight light lepton, a \loose but not \tight \tauhad with \(\pT > \qty{60}{\GeV}\) and a \tight \tauhad.
\end{itemize}

\begin{figure}[htbp]
\centering
\includegraphics[width=0.45\textwidth]{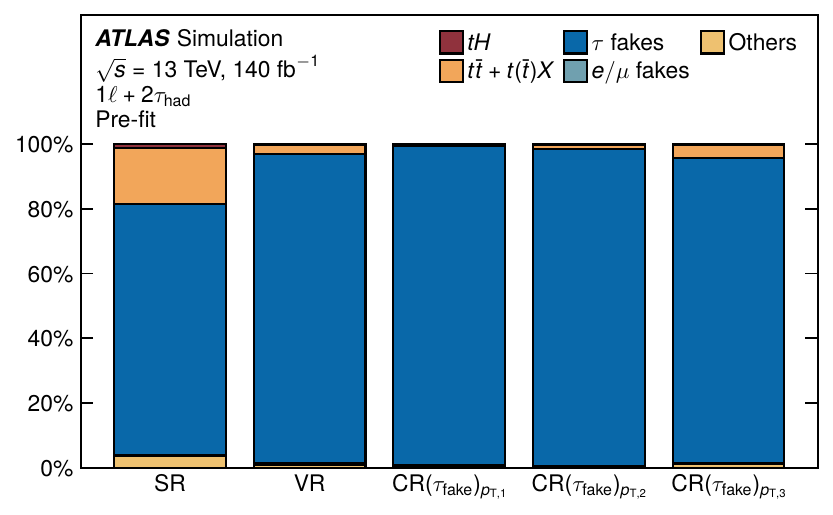}
\caption{Composition of the SR, the VR and the CRs used as input to the \lepditau profile likelihood fit.
\(\Pqt(\Paqt)X\) represents all processes, except the signal,
with one or two top quarks and a heavy boson (\PW, \PZ, \PH) in the final state.
}
\label{fig:fake_yields:lepditau}
\end{figure}

The composition of the different regions used in the fit is summarised in \cref{fig:fake_yields:lepditau}.
If the fraction of a fake process is less than one percent of all events in that region, it is not scaled in the fit.
Comparisons of the shapes of the light-lepton and fake-\tauhad kinematics in the SR, the CRs and the VR,
before the profile likelihood fit, show good agreement between data and simulation,
except for the \pT of the \tauhad fakes,
which motivates using three separate CRs with different \pT ranges.

%

%
%

%

%
\section{Systematic uncertainties}%
\label{sec:systematics}

The sources of systematic uncertainties considered and their treatment are in general identical
to those discussed in Ref.~\cite{HIGP-2024-03}.
They are grouped into three categories:
instrumental uncertainties that affect physics object reconstruction and identification;
uncertainties affecting the estimate of fake and non-prompt lepton backgrounds; and
modelling uncertainties relevant for MC simulation-based estimations. Their impact on the final fit is given in \cref{tab:mutH:MLbbgg:systDecomp}.

For the signal,
systematic uncertainties in the \tHq and \tWH MC predictions due to missing higher-order QCD corrections in the modelling
are estimated by doubling and halving the renormalisation and factorisation scales chosen for the nominal sample.
A systematic uncertainty due to the choice of the parton shower model is estimated
as the relative difference between the \POWHEG{}+\PYTHIA[8] and \POWHEG{}+\HERWIG[7]~\cite{Bahr:2008pv,Bellm:2015jjp} predictions and applied to the nominal predictions.
An uncertainty is assigned to the cross-section by summing in quadrature the contributions from QCD scale, flavour scheme, PDF and $\alpha_s$ from a NLO QCD prediction:
the values obtained are equal to \qty{15.4}{\%} and \qty{9.2}{\%} for \tHq and \tWH, respectively~\cite{deFlorian:2016spz,YR4twiki}.

Cross-section uncertainties in the background processes are estimated in a similar way.
Systematic uncertainties from different modelling choices are evaluated by comparing alternative simulated samples to the nominal ones.
These alternative samples were generated either through reweighting or with dedicated event generators.
Each variation is applied independently, unless explicitly stated otherwise,
to ensure a consistent evaluation of its individual impact on the analysis.
More details can be found in Ref.~\cite{HIGP-2024-03}.

As the analysis presented here makes use of \tauhad objects,
additional systematic uncertainties related to the reconstruction and identification of \tauhad are included~\cite{ATLAS-CONF-2017-029},
in both the \dileptau and \lepditau channels.
The uncertainties are derived from data and simulation comparisons using \(\PZ \to \Pgt\Pgt\) events.
These uncertainties consist of the following sources:

\begin{description}
\item[\tauhad reconstruction] derived by comparing the reconstruction efficiency
from \(\PZ \to \Pgt\Pgt\) \POWHEG{}+\PYTHIA[8] simulation and the simulations
using different \texttt{GEANT4} physics lists, alternative detector material simulations
and theory scale variations;
\item[\tauhad energy scale] derived from a \(\chi^2\) fit to data of the visible mass
of a \(\Pgm\)-\(\tauhad\) pair in a dedicated tag-and-probe analysis,
with additional uncertainties from the \(E/p\) measurement
and from usage of alternative detector material simulations, \texttt{GEANT4} physics lists
and theory scale variations;
\item[\tauhad RNN identification] derived from a dedicated \(\Pgm\)-\(\tauhad\)
tag-and-probe efficiency measurement;
\item[electron veto] (\textsc{eBDT} \texttt{loose} working point): derived from a dedicated \(e\)-\(\tauhad\) tag-and-probe efficiency measurement.
\end{description}

\subsection{Composition uncertainty -- \texorpdfstring{\tauhad}{tauhad} fakes}%
\label{sec:systematics:tau_comp}

Different objects can fake the signature of a \tauhad in the detector,
such as quark- or gluon-initiated jets, electrons or muons.
Possible differences in the composition between the data and MC simulation in the SRs and CRs are studied
using templates derived from MC simulation.
The templates are defined as follows:

\begin{description}
\item[tau] the reconstructed \tauhad is matched to a truth tau lepton but not to a truth jet;
\item[electron] the reconstructed \tauhad is matched to a truth electron but not to a truth jet;
\item[muon] the reconstructed \tauhad is matched to a truth muon but not to a truth jet;
\item[quark-initiated jet] the reconstructed \tauhad is matched to a jet generated by a quark, but not to a lepton;
\item[gluon-initiated jet] the reconstructed \tauhad is matched to a jet generated by a gluon, but not to a lepton;
\item[unknown] no match to a truth lepton or jet is found.
\end{description}

An uncertainty, determined using MC simulation, is added to both channels to cover the potential effect of differences
between the \tauhad-fake composition of the SR and the respective \pT-dependent CR.
The uncertainty is defined by comparing a template with the original SR composition with the one obtained by adjusting the relative distribution of quark-, gluon-, and light-lepton-initiated
\tauhad-fake events in a SR \pT bin to that of the corresponding CR.
As the overall normalisation of the SR is kept, the resulting uncertainty has a pure shape effect for \ONNtH.
In the case of the \lepditau channel, this uncertainty is applied independently for leading and sub-leading \tauhad-fake objects.
The uncertainty is evaluated by matching the composition in the SR to that in the CR for all three \tauhad(\pT) bins simultaneously.
The relative change in the composition of the NN bins in the SR is at most \qty{2.4}{\%} per bin.
\subsection{Composition uncertainty -- misidentified light leptons}
\label{sec:systematics:light_lepton_comp}

For the \dileptau channel, a composition uncertainty for the misidentified light leptons is introduced in the same way as for the \tauhad-fakes,
where the relative composition of the light-lepton misidentification sources in the SR is adjusted to match that in the respective CRs.
Due to the small contribution of leading misidentified light leptons, the uncertainty is determined only for sub-leading light leptons.
The scaling factors on the fractions of the different electron and muon fake sources
are determined separately for sub-leading electrons and muons, taking into account the different misidentification sources:
\begin{description}
\item[misidentified \Pe] heavy- and light-quark quark decays, \Pgt-leptons and photon conversions misidentified as electrons;
\item[misidentified \Pgm] heavy- and light-quark decays, and \Pgt-leptons misidentified as muons.
\end{description}
In the SR, the electron and muon factors are applied simultaneously to model the shape uncertainty
caused by the changed relative composition.
The relative change in the composition of the NN bins in the SR is typically \qty{2}{\%} per bin.


%

%
%

%

%
\section{Results}%
\label{sec:results}

The signal strength \mutH is determined from a maximum-likelihood fit,
where the best-fit value and the corresponding 68\% CL uncertainties are derived using the profile likelihood ratio
and its asymptotic distribution~\cite{Cowan:2010js}.
The \(\text{CL}_\text{s}\) method~\cite{Read:2002hq} is used to obtain the \qty{95}{\%} CL upper limit for \mutH,
assuming \(\mutH=0\) for the background-only hypothesis to derive the expected upper limit.
In the \mutH measurement, the total uncertainty is presented as a decomposition into separate components:
the \textit{statistical} component is obtained from a fit
in which the nuisance parameters associated with systematic uncertainties are fixed to their best-fit values;
the \textit{systematic} component, corresponding to the combined effect of systematic uncertainties,
is computed as the square root of the difference between the squares of the total uncertainty and the statistical uncertainty.

\begin{figure}[htbp]
\centering
\subcaptionbox{\label{fig:SR_SM:dilepSStau}}{%
\includegraphics[width=0.45\textwidth]{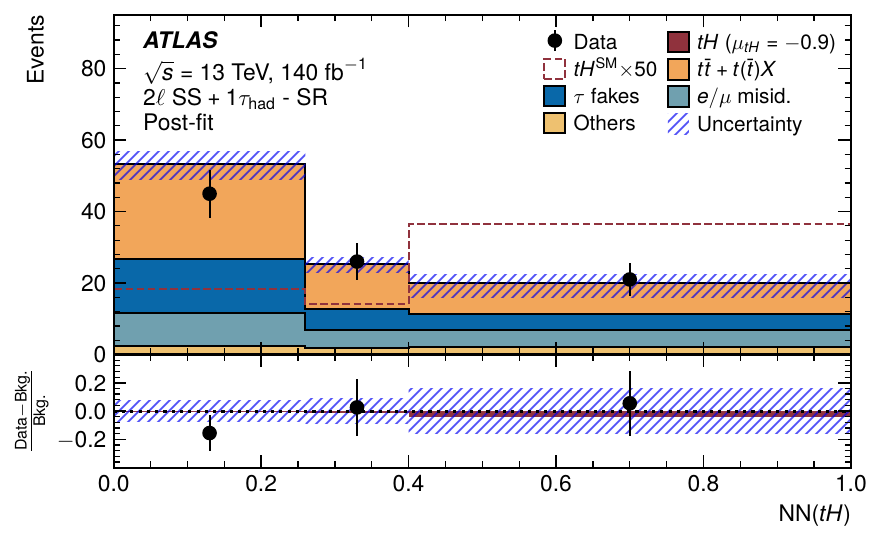}
}
\quad
\subcaptionbox{\label{fig:SR_SM:lepditau}}{%
\includegraphics[width=0.45\textwidth]{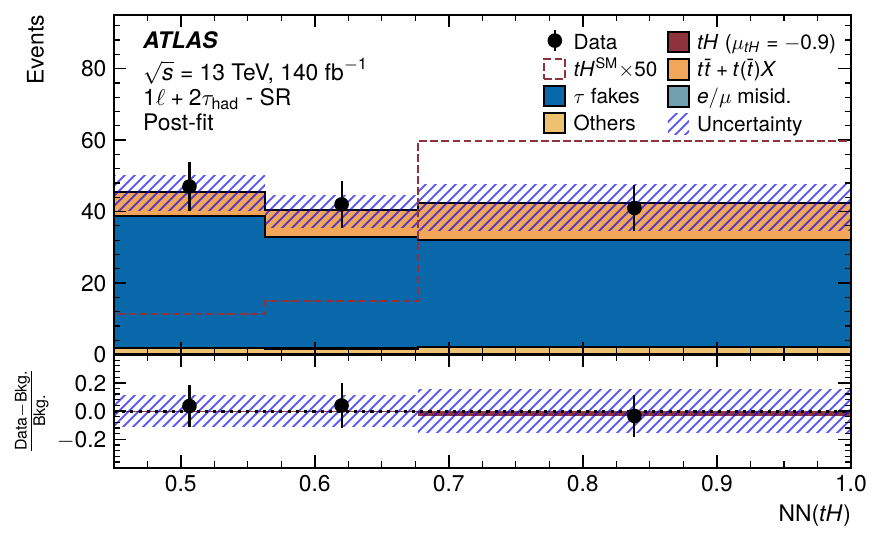}
}
\caption{Post-fit distributions of the \ONNtH score in the (a) \dilepSStau and (b) \lepditau SRs.
\(\Pqt(\Paqt)X\) represents all processes, except the signal,
with one or two top quarks and a heavy boson (\PW, \PZ, \PH) in the final state.
The lower panel shows the ratio of the post-fit signal to the background.
The (red) negative area in the lower panel corresponds to the post-fit signal yield being negative.
The uncertainties shown are the total uncertainties in the data and MC simulation.
}
\label{fig:SR_SM}
\end{figure}

\begin{figure}[htbp]
\centering
\subcaptionbox{\label{fig:CR_SM:dileptau}}{%
\includegraphics[width=0.45\textwidth]{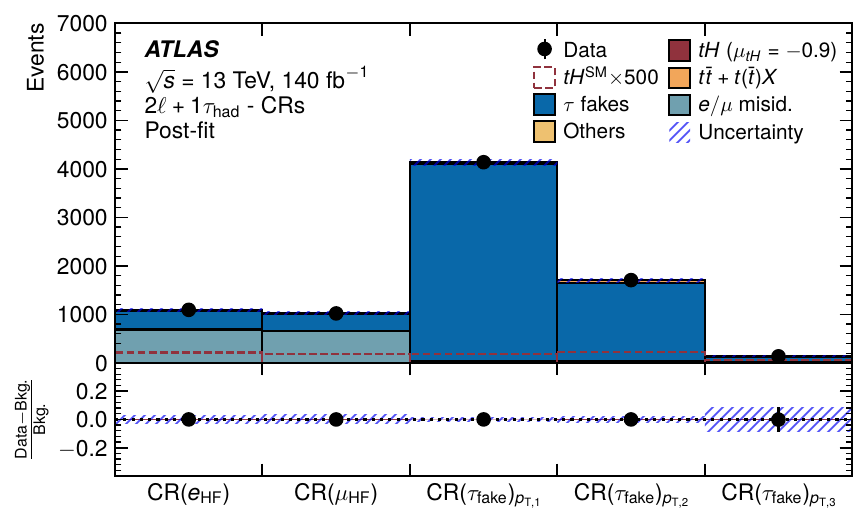}
}
\subcaptionbox{\label{fig:CR_SM:lepditau}}{%
\includegraphics[width=0.45\textwidth]{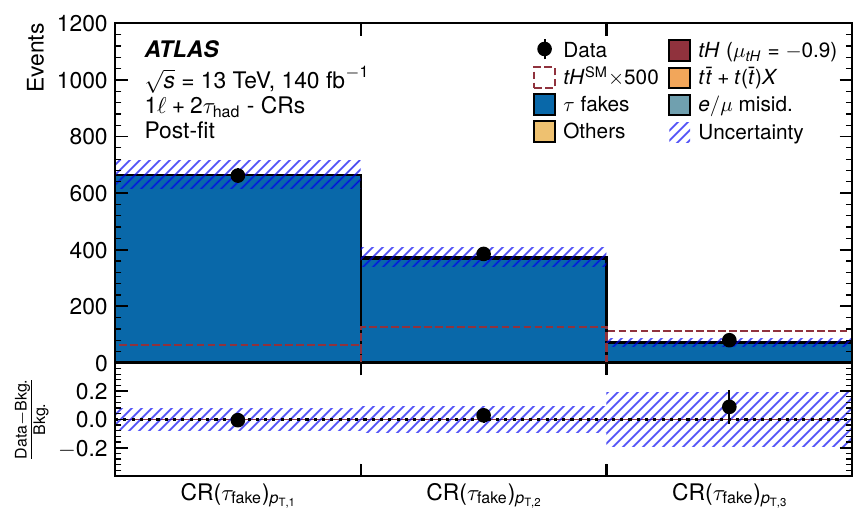}
}

\caption{Post-fit yields in the CRs for (a) the \dileptau and (b) the \lepditau channel.
\(\Pqt(\Paqt)X\) represents all processes, except the signal,
with one or two top quarks and a heavy boson (\PW, \PZ, \PH) in the final state.
The lower panel shows the ratio of the post-fit signal to the background.
The uncertainties shown are the total uncertainties in the data and MC simulation.
}
\label{fig:CR_SM}
\end{figure}

Separate fits in the \dileptau and \lepditau regions are conducted first,
using three \ONNtH bins in each of the SRs (see \cref{fig:SR_SM}),
in addition to the associated one-bin CRs listed in \cref{sec:regions}.
The \mutH is measured to be \(\mutHdileptauTot=\mutHdileptauStatSys\) in the \dileptau channel,
and \(\mutHlepditauTot=\mutHlepditauStatSys\) in the \lepditau channel.
A combined \mutH measured in the \Htautau channel is obtained
by fitting \dileptau and \lepditau regions simultaneously,
assuming that the \tauhad-fake scale factors are the same for these two regions,
and that the common experimental and theoretical uncertainties are correlated.
The \Htautau combined fit yields a signal strength of \(\mutHtautauTot=\mutHtautauStatSys\)
and a \qty{95}{\%} CL upper limit of \limtHtautauObs\,(\limtHtautauExp) times the SM prediction
in the observed\,(expected) case.
The observed and expected yields in the two SRs and the CRs
are shown in \cref{fig:SR_SM,fig:CR_SM}
and listed in \cref{sec:yields} \cref{tab:yields:SRCR:dileptau,tab:yields:SRCR:lepditau}.
The values of the NFs associated with the backgrounds are listed in \cref{tab:NF:bkg}.
Distributions of \ONNtH in the VRs are shown in \cref{sec:ONN:VRs}.
Good agreement between the data and the post-fit prediction is observed in all regions. The resulting \(\pt(\tauhad)\) distributions at pre-selection level are shown in \cref{fig:dilepSStau:NN:features:tau_pt,fig:lepditau:NN:features:tau_pt},
demonstrating good modelling of the fake background.

\begin{table}[htbp]
\tablesetup
\caption{Summary of the NFs scaling the background estimations
for the fits to the \dileptau and \lepditau channels and their combination.
The uncertainties include both statistical and systematic components.}
\label{tab:NF:bkg}

%
\sisetup{round-mode=places, round-precision=2}
\begin{tabular}{l *{3}{S[table-format = 1.2] @{\(\,\pm\,\)} S[table-format = 1.2]}}
\toprule
NF & \multicolumn{2}{c}{\dileptau} & \multicolumn{2}{c}{\lepditau} & \multicolumn{2}{c}{Combined}\\
\midrule
\ke      & 0.907196 & 0.0646472  & \multicolumn{2}{c}{--} & 0.899335 & 0.0594083 \\
\kmu     & 0.893541 & 0.0595     & \multicolumn{2}{c}{--} & 0.888102 & 0.0537082 \\
\ktau{1} & 1.09152  & 0.0528522  & 1.07516  & 0.0513503   & 1.09058  & 0.0349428 \\
\ktau{2} & 0.919856 & 0.0443811  & 0.945054 & 0.0681115   & 0.92203  & 0.0340759 \\
\ktau{3} & 0.626987 & 0.0807826  & 0.70947  & 0.207177    & 0.640298 & 0.0707427 \\
\bottomrule
\end{tabular}


\end{table}

Finally, the results in the \Htautau channel are combined
with the measurements from Ref.~\cite{HIGP-2024-03,HIGG-2020-16,HIGP-2024-24},
which target the same \tHchans processes,
but in final states in which the Higgs boson decays into \bbbar, light leptons or \gamgam.
The updated estimate of the fake-lepton contribution to the \Hbb analysis published in Ref.~\cite{HIGP-2024-03}
resulted in a small change to \mutH, which is now measured to be \(\mutHbbTot = \mutHbbStatSys\) in the \Hbb channel.
This new value and the updated uncertainty breakdown are used in the combinations discussed here.
Only the \ttH and \tH categories in the \Hgamgam analysis
from Ref.~\cite{HIGG-2020-16} are used in this combination.
The \ttW background yield measured in the multi-lepton channel
is used to control the \ttW background contribution in the \Htautau measurement,
by correlating the normalisation factors between these two channels.
In addition, the uncertainties in the reconstruction and calibration
of jets, muons and the \MET,
the uncertainties in the luminosity measurement and pile-up modelling from MC simulation,
and the theoretical uncertainties in \ttH, \tHq and \tWH from MC modelling are treated as correlated among all input channels using the full \Run{2} data set.
Between \Htautau, \Hbb and multi-lepton channels,
the electron reconstruction and calibration uncertainties,
the flavour-tagging uncertainties and the common theoretical uncertainties in background modelling are also correlated.
Between the \Run{2} and \Run{3} measurements in the \Hgg channel, a subset of uncertainties associated with the calibration of jets, missing transverse energy, and \(e/\gamma\) energy are treated as correlated, together with the uncertainties in the Higgs boson mass and the \Hgg branching-ratio.

\begin{figure}[htbp]
\centering
\subcaptionbox{}{%
\includegraphics[width=0.45\textwidth]{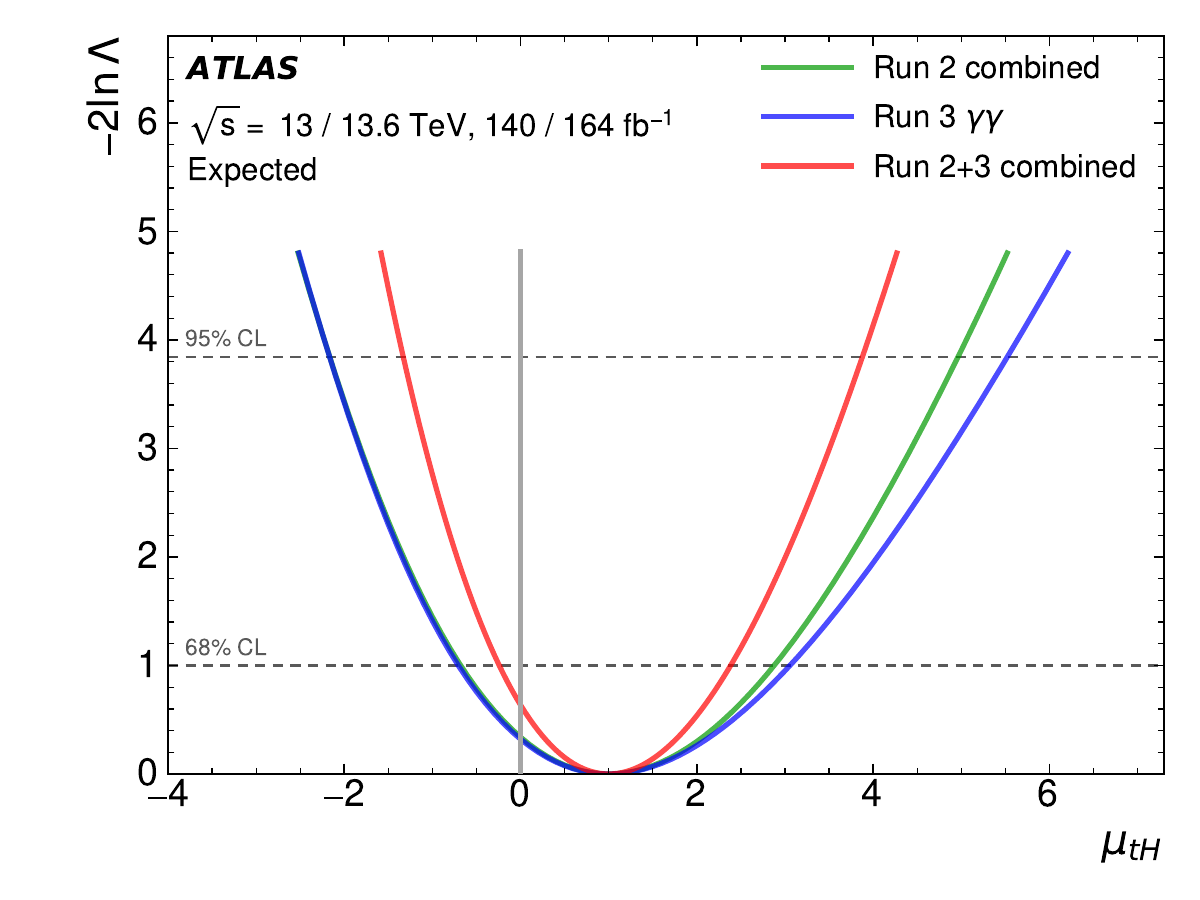}
}
\subcaptionbox{}{%
\includegraphics[width=0.45\textwidth]{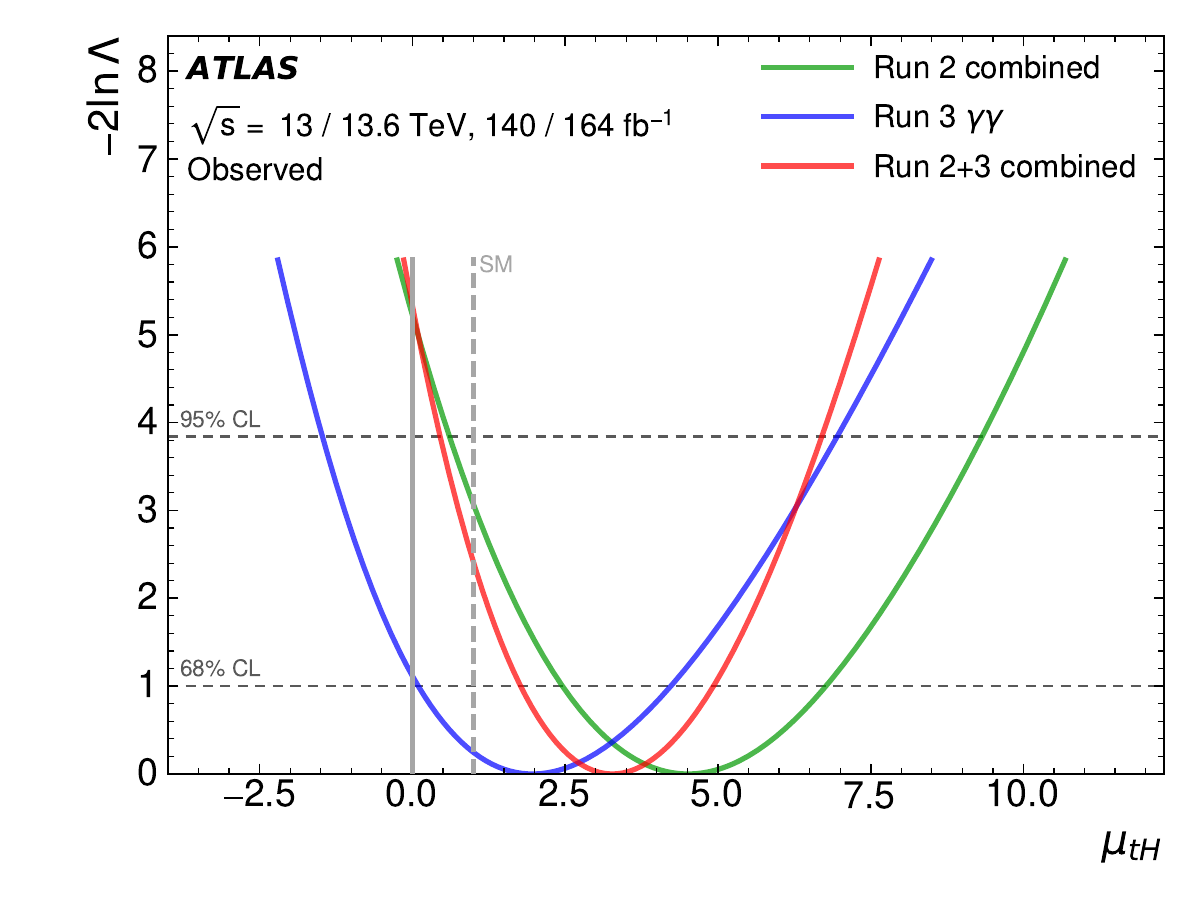}
}
\caption{Negative log-likelihood scans as a function of \mutH for the \Run{2} combined fit, the \Run{3} \Hgg fit, and the combined \Run{2}+\,\Run{3} fit in the (a) expected and (b) observed case.
The solid grey vertical line indicates the background-only hypothesis, corresponding to \(\mutH = 0\).
In the observed scan, the dashed grey vertical line indicates the SM prediction.}
\label{fig:mutH:MLbbtautaugg:NLLscan}
\end{figure}

\begin{figure}[htbp]
\centering
\includegraphics[width=0.8\textwidth]{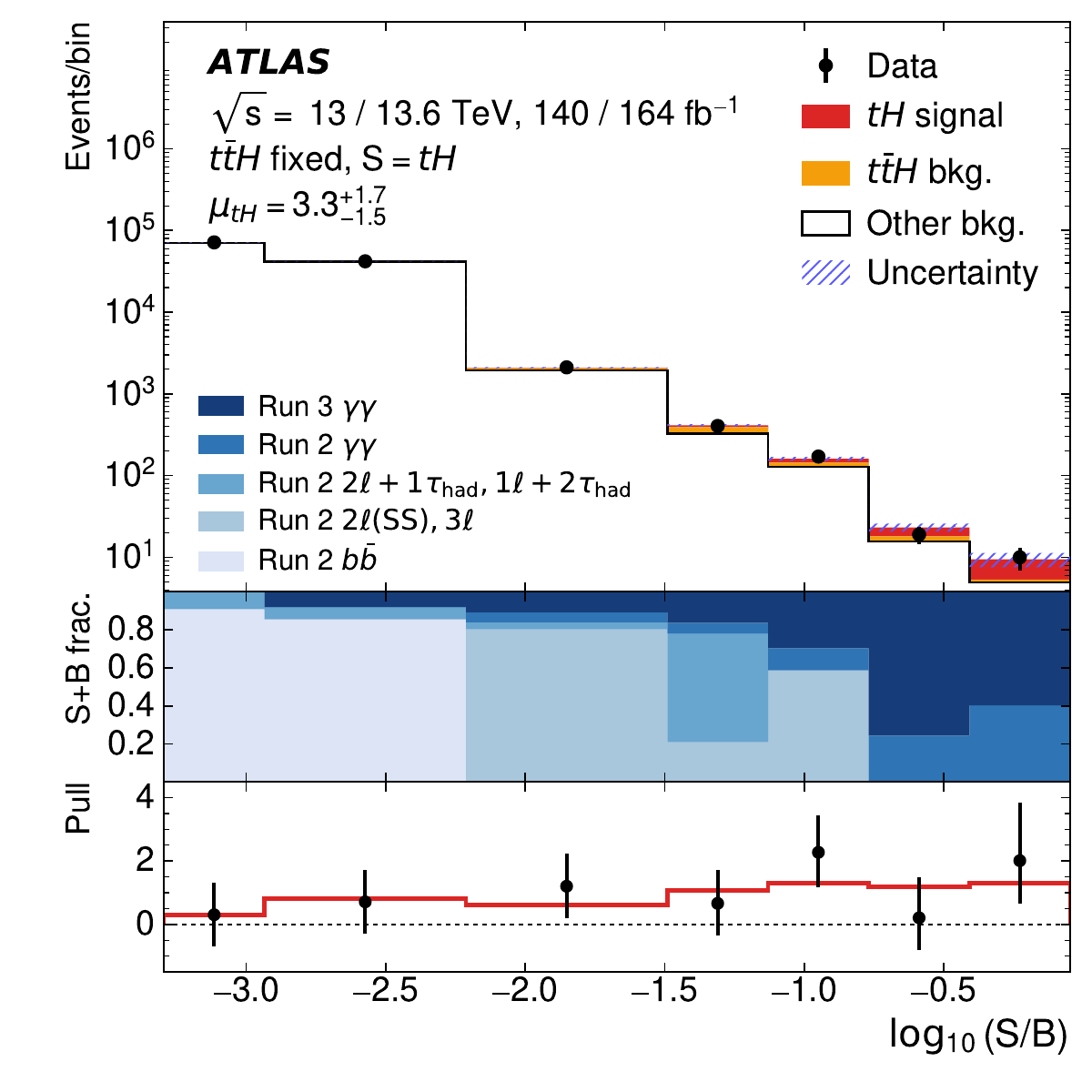}
\caption{Observed event yields in all analysis categories used in this combination.
The stacked histograms show the post-fit yields for \tH production, \ttH production, and all other background processes, obtained from the fit with \muttH fixed to its SM prediction.
For the \Hgg analysis, the yields are calculated only within the smallest \myy window containing 90\% of the \tH signal yield.
The discriminant bins in all categories are ranked by \(\log_{10}(S/B)\), where \(S\) is the \tH signal yield and \(B\) is the yield from all other processes, and are combined in decreasing order of \(\log_{10}(S/B)\).
The uncertainties shown on the simulated processes include both systematic and statistical sources.
The second panel shows the fraction of the total signal-plus-background (\(S+B\)) yield selected by each analysis.
The bottom panel shows the pull (the residual divided by the background uncertainty) of the data relative to the background-only prediction, where the uncertainty includes statistical contributions only.
The solid red line indicates the pull of the best-fit signal-plus-background prediction relative to the background-only prediction.}
\label{fig:mutH:MLbbtautaugg:sensitivityttHFixed}
\end{figure}

By fixing \muttH to the SM prediction,
the combined fit to all input channels yields a signal strength of
\begin{equation*}
\mutH = \mutHMLbbggMinosTot = \mutHMLbbggMinos\,.
\end{equation*}
The measured signal strength is compatible with the SM prediction at the 12\% level.
The expected and observed negative log-likelihood scans as a function of \mutH for the \Run{2} combined fit,
the \Run{3} \Hgg fit,
and the combined \Run{2}+\,\Run{3} fit are shown in \cref{fig:mutH:MLbbtautaugg:NLLscan}.
The combined fit achieves a significantly improved precision
compared with the \Run{2} combination or the \Run{3} \Hgg measurement.
The observed event yields and the relative contributions of the input measurements are summarised in \cref{fig:mutH:MLbbtautaugg:sensitivityttHFixed}.

\begin{figure}[htbp]
\centering
\includegraphics[width=0.8\textwidth]{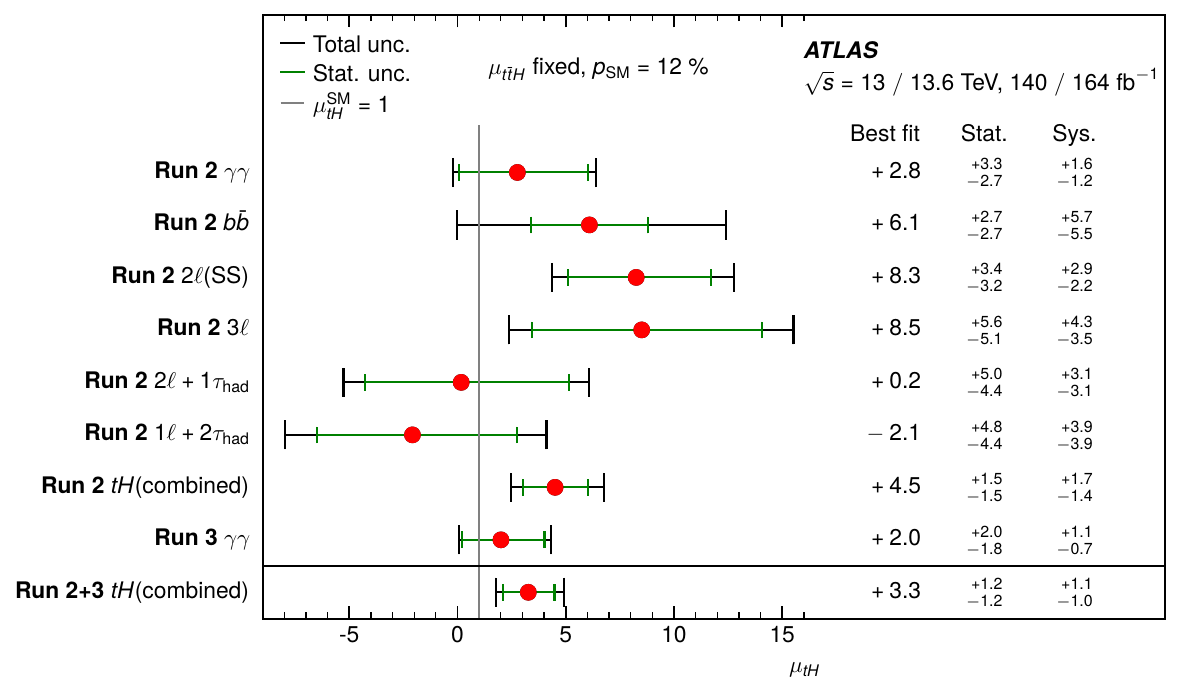}
\caption{Summary of the \mutH combined measurements. The individual \mutH in each channel
are obtained by floating them independently in the combined fit,
while \muttH is fixed to 1.
This approach accounts for nuisance-parameter correlations and
therefore differs from the individual results obtained in the standalone measurements.
The statistical uncertainty is obtained by fixing the nuisance parameters
associated with systematic uncertainties to their best-fit values.
The systematic uncertainty is computed as the square root of the difference
between the squares of the total uncertainty and the statistical uncertainty.
}
\label{fig:mutH:MLbbgg:fit}
\end{figure}

The best-fit values, total uncertainties and their decomposed components for \mutH are summarised in \cref{fig:mutH:MLbbgg:fit}
for the individual and combined results.
The individual results are obtained by allowing the \mutH of each input channel to float simultaneously in the fit to the full combination.
In the combined fit, the \mutH values in the \dileptau and \lepditau channels differ slightly from those
in their standalone measurements, because the \ttW yield is now constrained by the \ttW control regions
of the \dilepSS and \trilep analyses.
This change in the \ttW normalisation factor also contributes to the shift of \mutH in the \trilep channel,
together with the small effect from the slightly changed pulls of \ttbar and \diboson background modelling uncertainties in the combined fit,
compared with its standalone result reported in Ref.~\cite{HIGP-2024-03}.
Allowing each input channel to have an independent signal strength,
the compatibility with the combined result is found to correspond to a p-value of 68\%.

\begin{figure}[htbp]
\centering
\includegraphics[width=0.8\textwidth]{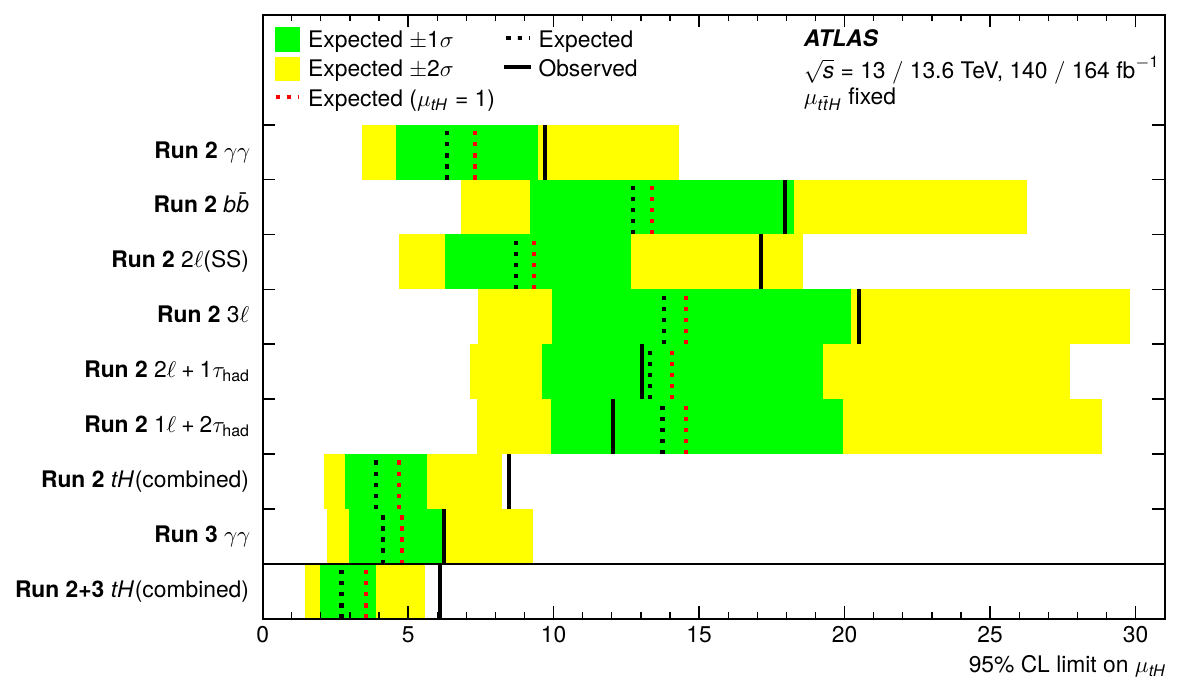}
\caption{Summary of the \qty{95}{\%} CL upper limit of \mutH
from the individual and combined analyses.
The error bands of the expected limit are obtained for \(\mutH=0\).
The dashed red line indicates the expected upper limit with \(\mutH = 1\).
}
\label{fig:mutH:MLbbgg:limit}
\end{figure}

A \qty{95}{\%} CL upper limit on the observed (expected) value of \mutH from the combined fit
is found to be \limtHMLbbggObs\,(\limtHMLbbggExp) times the SM prediction,
as shown in \cref{fig:mutH:MLbbgg:limit}.
The significance of \tH production is estimated to be \sigtHMLbbggObs\,(\sigtHMLbbggExp) standard deviations in the observed (expected) case.
By allowing \muttH to float in addition to \mutH,
the measured \tH signal strength is found to be
\(\mutHMLbbggttHFloatMinosTot = \mutHMLbbggttHFloatMinos\),
corresponding to a \qty{95}{\%} CL upper limit of \limtHMLbbggttHFloatObs\,(\limtHMLbbggttHFloatExp\,expected) times the SM prediction and an observed significance of \sigtHMLbbggttHFloatObs\,standard deviations.
These results are largely unchanged relative to those obtained with \muttH fixed to its SM value,
reflecting the strong constraint on \ttH production provided by the \Hgg measurement and the non-negligible \ttH contribution in the other \tH analyses.
In this fit configuration, \muttH is measured to be \muttHComb,
consistent with recent dedicated ATLAS measurements of \ttH production in the \Hgg, \Hbb and multi-lepton final states~\cite{HIGP-2024-24,HIGG-2020-24,HIGP-2024-08}.

An alternative decomposition of the total uncertainty into statistical and systematic components
can be performed using a BLUE-like method~\cite{Pinto:2023yob}
based on the post-fit covariance matrix,
assuming that the impact of each systematic uncertainty follows a Gaussian distribution.
Using this approach, the statistical and systematic components of the combined \mutH measurement are found to be 1.4 and 0.78, respectively.
This covariance-matrix-based method also provides a well-defined procedure
for further decomposing the systematic uncertainty into contributions from different sources.
The detailed breakdown of these grouped components is shown in \cref{tab:mutH:MLbbgg:systDecomp}.
The contributions from the experimental and theoretical uncertainties to the combined measurement are comparable in size.
On the experimental side, the dominant contribution arises from the jet energy calibration.
On the theoretical side, the largest contribution comes from signal modelling,
which includes missing higher-order QCD corrections, MC generator choices and parton shower modelling for the \tH process.

\begin{table}[htbp]
\tablesetup
\caption{Summary of the absolute values of the contributions from the main categories of uncertainties
on \mutH for the separate and the combined fits, calculated from the covariance matrix under the Gaussian distribution assumption.
The individual channel results are obtained from the individual fits to the input measurements of the final combination.
Some nuisance parameters with small impacts are grouped into categories, with the impact of the category
being the sum in quadrature of individual parameter impacts.
Categories that are not applicable to a given input measurement are indicated by a dash.
Indentation is used to denote a subcategory.
}
\resizebox{\textwidth}{!}{%

\begin{tabular}{lSSSSSSSSS}
\toprule
Source & \multicolumn{9}{c}{Impact on \mutH} \\
& \multicolumn{2}{c}{Combined} & \multicolumn{2}{c}{\Hgamgam} & \multicolumn{1}{c}{\Hbb} & \multicolumn{1}{c}{\dilepSS} & \multicolumn{1}{c}{\trilep} & \multicolumn{1}{c}{\dileptau} & \multicolumn{1}{c}{\lepditau} \\
& \multicolumn{1}{c}{All} & \multicolumn{1}{c}{\Run{2}} & \multicolumn{1}{c}{\Run{3}} & \multicolumn{1}{c}{\Run{2}} & \multicolumn{1}{c}{\Run{2}} & \multicolumn{1}{c}{\Run{2}} & \multicolumn{1}{c}{\Run{2}} & \multicolumn{1}{c}{\Run{2}} & \multicolumn{1}{c}{\Run{2}} \\

\midrule
\textbf{Experimental uncertainties} & 0.53 & 0.76 & 0.43 & 1.1 & 1.5 & 1.6 & 2.6 & 1.5 & 1.5 \\
\quad JES and JER & 0.40 & 0.59 & 0.31 & 0.17 & 0.63 & 1.6 & 2.5 & 1.2 & 1.2 \\
\quad \(e/\gamma\) & 0.20 & 0.27 & 0.12 & 0.85 & {--} & 0.01 & 0.07 & 0.06 & 0.02 \\
\quad Photons & 0.14 & 0.09 & 0.20 & 0.22 & {--} & {--} & {--} & {--} & {--} \\
\quad Spurious signal & 0.12 & 0.16 & 0.12 & 0.51 & {--} & {--} & {--} & {--} & {--} \\
\quad Taus & 0.11 & 0.20 & {--} & {--} & {--} & {--} & {--} & 0.63 & 0.82 \\
\quad Fake-related & 0.11 & 0.19 & {--} & {--} & 1.2 & 0.22 & 0.08 & 0.42 & 0.14 \\
\quad Pile-up & 0.06 & 0.12 & 0.05 & 0.09 & 0.05 & 0.05 & 0.45 & 0.10 & 0.17 \\
\quad Flavour-tagging & 0.05 & 0.06 & 0.05 & 0.06 & 0.64 & 0.13 & 0.16 & 0.26 & 0.24 \\
\quad Electrons & 0.04 & 0.03 & 0.05 & 0.01 & 0.09 & 0.11 & 0.30 & 0.22 & 0.00 \\
\quad Muons & 0.03 & 0.04 & 0.01 & 0.01 & 0.00 & 0.12 & 0.40 & 0.23 & 0.05 \\
\quad \MET & 0.03 & 0.06 & 0.00 & 0.05 & 0.04 & 0.20 & 0.03 & 0.03 & 0.08 \\
\quad JVT and fJVT & 0.01 & 0.04 & {--} & 0.06 & 0.16 & 0.11 & 0.06 & 0.02 & 0.10 \\
\quad Other experimental & 0.08 & 0.13 & 0.10 & 0.39 & 0.10 & 0.12 & 0.14 & 0.23 & 0.07 \\
\midrule
\textbf{Theory uncertainties} & 0.54 & 0.77 & 0.52 & 0.57 & 3.4 & 1.4 & 2.5 & 2.8 & 2.5 \\
\quad Signal & 0.42 & 0.46 & 0.52 & 0.57 & 0.69 & 0.97 & 1.0 & 0.46 & 0.44 \\
\quad \ttb & 0.19 & 0.34 & {--} & {--} & 3.0 & {--} & {--} & {--} & {--} \\
\quad \ttc and \ttlight & 0.17 & 0.28 & {--} & {--} & 1.3 & {--} & {--} & {--} & {--} \\
\quad \ttbar & 0.14 & 0.24 & {--} & {--} & {--} & 0.52 & 1.9 & 1.6 & 1.9 \\
\quad \Diboson & 0.13 & 0.24 & {--} & {--} & 0.06 & 0.22 & 1.0 & 1.7 & 0.91 \\
\quad \ttZ & 0.05 & 0.09 & {--} & {--} & 0.05 & 0.07 & 0.47 & 0.99 & 0.19 \\
\quad \ttW & 0.05 & 0.08 & {--} & {--} & 0.00 & 0.52 & 0.18 & 0.83 & 0.15 \\
\quad Other background & 0.13 & 0.24 & {--} & {--} & 0.60 & 0.58 & 0.43 & 0.56 & 1.3 \\
\midrule
\textbf{MC simulation statistics} & 0.20 & 0.37 & {--} & {--} & {--} & 1.1 & 1.7 & 0.97 & 1.5 \\
\midrule
\textbf{Total systematic uncertainty} & 0.78 & 1.1 & 0.67 & 1.2 & 3.7 & 2.4 & 4.0 & 3.3 & 3.3 \\
\textbf{Statistical uncertainty} & 1.4 & 1.8 & 1.9 & 3.0 & 4.9 & 3.5 & 5.6 & 4.8 & 5.2 \\
\textbf{Total} & 1.6 & 2.1 & 2.0 & 3.3 & 6.2 & 4.2 & 6.9 & 5.9 & 6.1 \\
\bottomrule
\end{tabular}


%
}
\label{tab:mutH:MLbbgg:systDecomp}
\end{table}

%
%
%
%
%
%
%
%

%
%
%
%
%
%
%

%
%
%
%
%
%
%
%


%

%
%
%
%

%
%
\FloatBarrier

%
%

%

%
\section{Conclusion}%
\label{sec:conclusion}

A search for the production process \(\Pp\Pp \to \tH\) is presented,
by selecting events with signatures corresponding to
the decay of the Higgs boson to \Pgt-leptons,
where at least one of the \Pgt-leptons decays hadronically.
The analysis is based on \pp collision data
at a centre-of-mass energy of \(\sqrt{s} = \qty{13}{\TeV}\)
recorded by the ATLAS detector at the LHC
from 2015 to 2018, corresponding to an integrated luminosity of \lumi.
The signal regions are split into final states with one or two \tauhad,
with additional control regions to constrain the main backgrounds.
The signal strength is found to be
\(\mutH = \mutHtautauTot\,,\)
leading to a \qty{95}{\%} CL upper limit on the observed (expected) value of \(\mu_{\tH}\)
of \limtHtautauObs\,(\limtHtautauExp) times the SM prediction.

Combining these measurements with those published in Refs.~\cite{HIGP-2024-03,HIGG-2020-16,HIGP-2024-24}
leads to a signal strength of
\(\mutH = \mutHMLbbggMinos\),
with a \qty{95}{\%} CL upper limit on the observed\,(expected) value of \(\mu_{\tH}\)
of \limtHMLbbggObs\,(\limtHMLbbggExp) times the SM prediction.
The measured signal strength is compatible with the SM prediction at the 12\% level.
The significance of \tH production is found to be \sigtHMLbbggObs\,(\sigtHMLbbggExp) standard deviations in the observed (expected) case.
This combination constitutes the most precise measurement of \tH production to date.


%

%
%
%
%
%
\section*{Acknowledgements}
%

%

%
%

%
%

We thank CERN for the very successful operation of the LHC and its injectors, as well as the support staff at
CERN and at our institutions worldwide without whom ATLAS could not be operated efficiently.

The crucial computing support from all WLCG partners is acknowledged gratefully, in particular from CERN, the ATLAS Tier-1 facilities at TRIUMF/SFU (Canada), NDGF (Denmark, Norway, Sweden), CC-IN2P3 (France), KIT/GridKA (Germany), INFN-CNAF (Italy), NL-T1 (Netherlands), PIC (Spain), RAL (UK) and BNL (USA), the Tier-2 facilities worldwide and large non-WLCG resource providers. Major contributors of computing resources are listed in Ref.~\cite{ATL-SOFT-PUB-2026-001}.

We gratefully acknowledge the support of ANPCyT, Argentina; YerPhI, Armenia; ARC, Australia; BMWFW and FWF, Austria; ANAS, Azerbaijan; CNPq and FAPESP, Brazil; NSERC, NRC and CFI, Canada; CERN; ANID, Chile; CAS, MOST, NSFC and FRFCU, China; Minciencias, Colombia; MEYS CR, Czech Republic; DNRF and DNSRC, Denmark; IN2P3-CNRS and CEA-DRF/IRFU, France; SRNSFG, Georgia; BMFTR, HGF and MPG, Germany; GSRI, Greece; RGC and Hong Kong SAR, China; ICHEP and Academy of Sciences and Humanities, Israel; INFN, Italy; MEXT and JSPS, Japan; CNRST, Morocco; NWO, Netherlands; RCN, Norway; MNiSW, Poland; FCT, Portugal; MNE/IFA, Romania; MSTDI, Serbia; MSSR, Slovakia; ARIS and MVZI, Slovenia; DSI/NRF, South Africa; MICIU/AEI, Spain; SRC and Wallenberg Foundation, Sweden; SERI, SNSF and Cantons of Bern and Geneva, Switzerland; NSTC, Taipei; TENMAK, T\"urkiye; STFC/UKRI, United Kingdom; DOE and NSF, United States of America.

Individual groups and members have received support from BCKDF, CANARIE, CRC and DRAC, Canada; CERN-CZ and PRIMUS, Czech Republic; COST, ERC, ERDF, Horizon 2020 and Marie Sk{\l}odowska-Curie Actions, European Union; Investissements d'Avenir Labex, Investissements d'Avenir Idex and ANR, France; DFG and AvH Foundation, Germany; Herakleitos, Thales and Aristeia programmes co-financed by EU-ESF and the Greek NSRF, Greece; BSF-NSF and MINERVA, Israel; NCN and NAWA, Poland; La Caixa Banking Foundation, CERCA and AGAUR programs from Generalitat de Catalunya and PROMETEO and GenT Programmes Generalitat Valenciana, Spain; G\"{o}ran Gustafssons Stiftelse, Sweden; The Royal Society and Leverhulme Trust, United Kingdom; Eric and Wendy Schmidt Fund for Strategic Innovation, United States of America.

In addition, individual members wish to acknowledge support from Chile: Agencia Nacional de Investigaci\'on y Desarrollo (ANID FONDECYT reg. 1230987, FONDECYT 1230812, FONDECYT 1240864, Fondecyt 3240661, Fondecyt Regular 1240721); China: Fundamental Research Funds for the Central Universities (010-63263105), Chinese Ministry of Science and Technology (MOST-2023YFA1605700, MOST-2023YFA1609300), National Natural Science Foundation of China (NSFC 12275265, NSFC-W2543005); Czech Republic: Czech Science Foundation (GACR - 24-11373S), Ministry of Education Youth and Sports (ERC-CZ-LL2327, FORTE CZ.02.01.01/00/22\_008/0004632); EU: H2020 European Research Council (ERC - 101002463); European Union: European Research Council (BARD No. 101116429, ERC - 101219398, ERC 101089007), European Regional Development Fund (HE COFUND GA No.101081355, ERDF), Marie Sklodowska-Curie Actions (GAP-101168829); France: Agence Nationale de la Recherche (ANR-22-EDIR-0002, ANR-24-CE31-0504-01); Germany: Deutsche Forschungsgemeinschaft (DFG - 469666862); China: Research Grants Council (GRF); Italy: Istituto Nazionale di Fisica Nucleare (LHC-MIUR - 28003/2025), Ministero dell'Università e della Ricerca (NextGenEU I53D23000820006 M4C2.1.1, SOE2024\_0000023); Japan: Japan Society for the Promotion of Science (JSPS KAKENHI  JP25H0063, JSPS KAKENHI JP22H04944, JSPS KAKENHI JP24K23939, JSPS KAKENHI JP24KK0251, JSPS KAKENHI JP25H00650, JSPS KAKENHI JP25H01291, JSPS KAKENHI JP25K01011, JSPS KAKENHI JP25K01023, JSPS KAKENHI JP25KK0047); Poland: Polish National Science Centre (NCN 2021/42/E/ST2/00350, NCN OPUS 2023/51/B/ST2/02507, NCN OPUS nr 2022/47/B/ST2/03059, NCN UMO-2019/34/E/ST2/00393, UMO-2022/47/O/ST2/00148, UMO-2023/49/B/ST2/04085, UMO-2023/51/B/ST2/00920, UMO-2024/53/N/ST2/00869); Spain: Agència de Gestió d'Ajuts Universitaris i de Recerca. (AGAUR - 2023 BP 00141), Ministry of Science and Innovation (RYC2019-028510-I, RYC2020-030254-I, RYC2021-031273-I, RYC2022-038164-I), Ministerio de Ciencia, Innovación y Universidades/Agencia Estatal de Investigaci\'on (EU NextGenerationEU (PRTR-C17.I1), PID2022-142604OB-C22); Sweden: Carl Trygger Foundation (Carl Trygger Foundation CTS 22:2312), Swedish Research Council (Swedish Research Council 2023-04654, VR 2021-03651, VR 2022-03845, VR 2022-04683, VR 2023-03403, VR 2024-05451, VR 2025-05940), Knut and Alice Wallenberg Foundation (KAW 2023.0366); United Kingdom: The Binks Trust, Royal Society (NIF-R1-231091); United States of America: U.S. Department of Energy (ECA DE-AC02-76SF00515), John Templeton Foundation (John Templeton Foundation 63206), Neubauer Family Foundation.

%
%


%
\clearpage
\appendix
\part*{Appendix}
\addcontentsline{toc}{part}{Appendix}
\crefalias{section}{appendix}

%

%
%
%
\section{Features used in the neural networks}%
\label{sec:NN-features}

The features used in the training of the NNs are listed in \cref{tab:NN-features:dileptau,tab:NN-features:lepditau}.

\begin{table}[htbp]
\tablesetup
\caption{Features used in the NN training in the \dilepSStau channel.}
\label{tab:NN-features:dileptau}
\begin{tabular}{l}
\toprule
\(\pT, \eta, \phi\) of the leading jet \\
\(\pT, \eta, \phi\) of the spectator jet \\
\(\pT, \eta, \phi\) of the leading \bjet \\
\(\pT, \eta, \phi\) of the light lepton associated with the top quark \\
\(\pT, \eta, \phi\) of the light lepton associated with the Higgs boson \\
\(\pT, \eta, \phi\) of the \tauhad \\
Missing transverse energy, \MET \\
Sum of the transverse energies of visible particles \\
\pT of the visible top-quark decay products \\
\(\eta\) of the visible top-quark decay products \\
Mass of the Lorentz-vector sum of the visible top-quark decay products \\
\pT of the Lorentz-vector sum of the \tauhad and the associated light lepton \\
\(\eta\) of the Lorentz-vector sum of the \tauhad and the associated light lepton \\
\DeltaR of the \tauhad and the associated light lepton \\
\(\Delta\phi\) of the \tauhad and the associated light lepton \\
Reconstructed mass of the Higgs boson using the MMC \\
Mass of the Lorentz-vector sum of the leading \bjet and spectator jet \\
\bottomrule
\end{tabular}
\end{table}

\begin{table}[htbp]
\tablesetup
\caption{Features used in the NN training in the \lepditau channel.}
\label{tab:NN-features:lepditau}
\centering
\begin{tabular}{l}
\toprule
\(\pT, \eta, \phi\) of the leading jet \\
\(\pT, \eta, \phi\) of the spectator jet \\
\(\pT, \eta, \phi\) of the leading \bjet \\
\(E, \pT, \eta, \phi\) of the light lepton associated with the top quark \\
\(E, \pT, \eta, \phi\) of the leading \tauhad \\
\(E, \pT, \eta, \phi\) of the sub-leading \tauhad \\
Missing transverse energy, \MET \\
Sum of the transverse energies of the visible particles \\
\pT of the visible top-quark decay products \\
\(\eta\) of the visible top-quark decay products \\
\pT of the Lorentz-vector sum of the two \tauhad \\
\(\eta\) of the Lorentz-vector sum of the two \tauhad \\
\DeltaR of the two \tauhad \\
\(\Delta\phi\) of the two \tauhad \\
Reconstructed mass of the Higgs boson using the MMC \\
Mass of the Lorentz-vector sum of the leading \bjet and spectator jet \\
\bottomrule
\end{tabular}
\end{table}


\clearpage

\section{Post-fit yields}%
\label{sec:yields}

The post-fit yields in the signal and control regions for both channels are given
in \cref{tab:yields:SRCR:dileptau,tab:yields:SRCR:lepditau}.

%

\begin{table}[htbp]
\tablesetup
\sisetup{table-align-uncertainty}%
\caption{Observed post-fit event yields in the \dileptau channel,
after the simultaneous fit to the \dileptau and \lepditau regions.
The contributions from single top (\tW), \Zjets and \Wjets are negligible in all regions and are therefore not shown.
Negligible contributions in a single region are shown as \enquote{{--}}.
The uncertainties correspond to the total uncertainties in the simulated samples.
}
\label{tab:yields:SRCR:dileptau}
\resizebox{\textwidth}{!}{
\begin{tabular}{l |
*{6}{S[table-format = 4.2(4), separate-uncertainty]}
}
\toprule
Channel & \multicolumn{6}{c}{Event yields} \\
& {\SR} & {\CRe} & {\CRmu} & {\CRtau{1}} & {\CRtau{2}} & {\CRtau{3}} \\
\midrule
\tHq     & -1 \pm 4 & -0.3 \pm 1.2 & -0.3 \pm 1.0 & -0.2 \pm 0.9 & -0.2 \pm 0.8 & -0.03 \pm 0.11 \\
\tWH     & -0.5 \pm 2.0 & -0.2 \pm 0.7 & -0.1 \pm 0.5 & -0.2 \pm 0.8 & -0.3 \pm 1.1 & -0.1 \pm 0.4 \\
\tWZ     & 1.4 \pm 0.7 & 0.42 \pm 0.22 & 0.31 \pm 0.16 & 1.7 \pm 0.9 & 2.6 \pm 1.3 & 0.9 \pm 0.5 \\
\ttbar   & 0.25 \pm 0.03 & 1.51 \pm 0.16 & 0.60 \pm 0.05 & 0.11 \pm 0.02 & 0.15 \pm 0.04 & {--} \\
\ttW     & 20.1 \pm 3.0 & 3.0 \pm 0.4 & 2.9 \pm 0.4 & 11.5 \pm 1.9 & 16.2 \pm 2.9 & 5.6 \pm 1.8 \\
\ttZ     & 9.2 \pm 3.2 & 2.7 \pm 0.8 & 2.4 \pm 1.2 & 12.8 \pm 3.0 & 17 \pm 4 & 5.1 \pm 1.1 \\
\ttH     & 13.6 \pm 1.6 & 4.3 \pm 0.6 & 3.5 \pm 0.4 & 6.3 \pm 0.7 & 9.5 \pm 1.0 & 2.74 \pm 0.35 \\
\tZq     & 3.18 \pm 0.34 & 0.99 \pm 0.13 & 0.89 \pm 0.14 & 5.2 \pm 0.5 & 5.4 \pm 0.5 & 0.83 \pm 0.09 \\
\Diboson              & 5.8 \pm 2.1 & 1.8 \pm 0.5 & 1.5 \pm 0.5 & 8.4 \pm 2.6 & 11 \pm 4 & 3.8 \pm 1.2 \\
Other Higgs           & {--} & {--} & {--} & 0.73 \pm 0.26 & {--} & {--} \\
Minor bkgs.           & 0.63 \pm 0.33 & 0.18 \pm 0.09 & 0.5 \pm 0.5 & 0.28 \pm 0.14 & 0.43 \pm 0.22 & 0.16 \pm 0.09 \\
Other fakes           & 4.5 \pm 1.4 & 10.1 \pm 0.9 & 23.2 \pm 1.4 & 26.3 \pm 1.6 & 16.9 \pm 0.7 & 2.5 \pm 0.5 \\
\Pgt fakes            & 18.4 \pm 2.7 & 5.4 \pm 0.7 & 4.0 \pm 0.9 & 4060 \pm 60 & 1630 \pm 40 & 113 \pm 12 \\
\Pe fakes             & 7.7 \pm 2.0 & 637 \pm 29 & 36 \pm 8 & {--} & {--} & {--} \\
\Pgm fakes            & 6.6 \pm 3.2 & 39 \pm 8 & 590 \pm 40 & {--} & {--} & {--} \\
\(\Pgt + \Pe\) fakes  & 3.9 \pm 1.4 & 363 \pm 21 & 24 \pm 4 & {--} & {--} & {--} \\
\(\Pgt + \Pgm\) fakes & 3.3 \pm 1.0 & 24.3 \pm 2.3 & 324 \pm 31 & {--} & {--} & {--} \\
\midrule
Total  & 97 \pm 7 & 1092 \pm 33 & 1019 \pm 32 & 4140 \pm 60 & 1710 \pm 40 & 134 \pm 12 \\
\midrule
Data   & 92 & 1093 & 1019 & 4137 & 1706 & 134 \\
\bottomrule
\end{tabular}
}
\end{table}

\begin{table}[htbp]
\tablesetup
\sisetup{table-align-uncertainty}%
\caption{Observed post-fit event yields in the \lepditau channel,
after the simultaneous fit to the \dileptau and \lepditau regions.
Processes with a negligible number of events are omitted (\ttbar, \Zjets, \Wjets and single top (\(t\)-ch., \tW)).
Negligible contributions in a single region are shown as \enquote{{--}}.
The uncertainties correspond to the total uncertainties in the simulated samples.
}
\label{tab:yields:SRCR:lepditau}
\begin{tabular}{l |
*{4}{S[table-format = 4.2(4), separate-uncertainty]}
}
\toprule
Channel & \multicolumn{4}{c}{Event yields} \\
& {\SR} & {\CRtau{1}} & {\CRtau{2}} & {\CRtau{3}} \\
\midrule
\tHq         & -2 \pm 6 & -0.1 \pm 0.4 & -0.2 \pm 0.8 & -0.2 \pm 0.7 \\
\tWH         & -0.5 \pm 1.8 & -0.04 \pm 0.15 & -0.06 \pm 0.23 & -0.07 \pm 0.26 \\
\tWZ         & 1.2 \pm 0.6 & 0.14 \pm 0.07 & 0.27 \pm 0.14 & 0.22 \pm 0.20 \\
\ttW         & 1.0 \pm 0.8 & 0.38 \pm 0.14 & 0.58 \pm 0.12 & 0.41 \pm 0.11 \\
\ttZ         & 7.6 \pm 1.1 & 0.74 \pm 0.15 & 1.45 \pm 0.29 & 1.13 \pm 0.19 \\
\ttH         & 9.9 \pm 1.4 & 0.96 \pm 0.20 & 1.89 \pm 0.30 & 1.6 \pm 0.4 \\
\tZq         & 4.9 \pm 0.6 & 0.64 \pm 0.09 & 0.99 \pm 0.11 & 0.67 \pm 0.10 \\
\Diboson     & 5.5 \pm 3.3 & 1.4 \pm 0.9 & 1.4 \pm 0.7 & 1.1 \pm 0.6 \\
Minor bkgs.  & 0.05 \pm 0.03 & {--} & 0.01 \pm 0.01 & 0.01 \pm 0.01 \\
Other fakes  & 0.31 \pm 0.07 & 2.7 \pm 0.4 & 1.1 \pm 0.4 & 0.25 \pm 0.03 \\
\Pgt fake    & 67 \pm 13 & 303 \pm 20 & 155 \pm 12 & 35 \pm 5 \\
Di-\Pgt fake & 30 \pm 4 & 355 \pm 34 & 211 \pm 29 & 33 \pm 14 \\
\midrule
Total  & 126 \pm 14 & 670 \pm 50 & 374 \pm 34 & 73 \pm 14 \\
\midrule
Data   & 130 & 662 & 385 & 80 \\
\bottomrule
\end{tabular}
\end{table}


\clearpage

\section{Neural network outputs in the validation regions}%
\label{sec:ONN:VRs}

\begin{figure}[htbp]
\centering
\subcaptionbox{\label{fig:VR_SM:dilepSStau}}{%
\includegraphics[width=\figwidth]{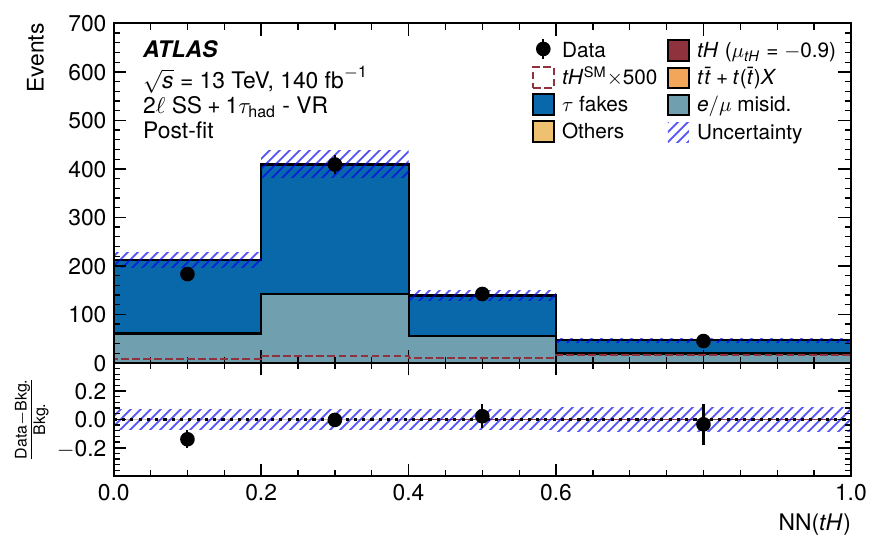}
}
\quad
\subcaptionbox{\label{fig:VR_SM:lepditau}}{%
\includegraphics[width=\figwidth]{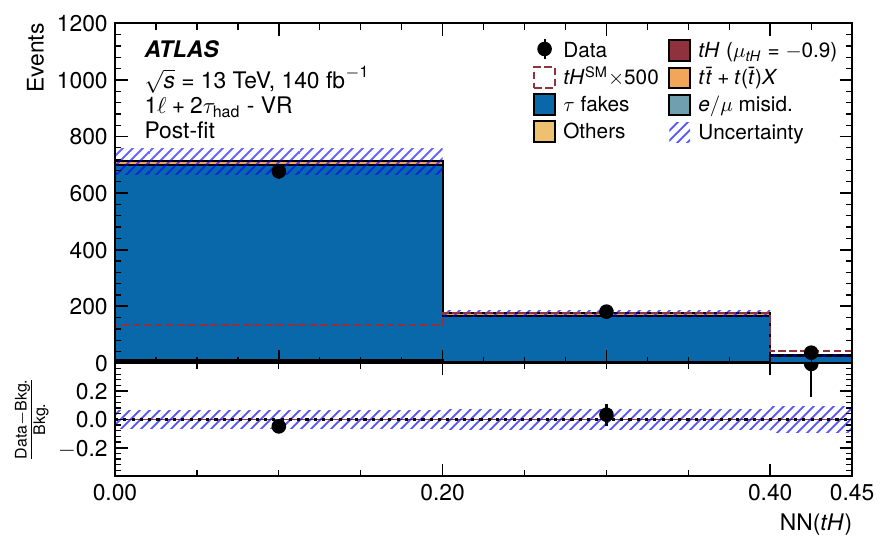}
}
\caption{Distributions of \ONNtH in the VR normalised to the post-fit yields.
(a) \dilepSStau and (b) \lepditau VRs.
\(\Pqt(\Paqt)X\) represents all processes, except the signal,
with one or two top quarks and a heavy boson (\PW, \PZ, \PH) in the final state.
The lower panel shows the ratio of the post-fit signal to the background.
The uncertainties shown are the total uncertainties in the data and MC simulation.
}
\label{fig:VR_SM}
\end{figure}

The distributions of \ONNtH in the VRs are shown in \cref{fig:VR_SM}.

\clearpage


%

%
%
\printbibliography
\clearpage
 
\begin{flushleft}
\hypersetup{urlcolor=black}
{\Large The ATLAS Collaboration}

\bigskip

\AtlasOrcid[0000-0002-6665-4934]{G.~Aad}$^\textrm{\scriptsize 102}$,
\AtlasOrcid[0000-0001-7616-1554]{E.~Aakvaag}$^\textrm{\scriptsize 17}$,
\AtlasOrcid[0000-0002-5888-2734]{B.~Abbott}$^\textrm{\scriptsize 121}$,
\AtlasOrcid[0000-0002-0287-5869]{S.~Abdelhameed}$^\textrm{\scriptsize 83b}$,
\AtlasOrcid[0000-0002-1002-1652]{K.~Abeling}$^\textrm{\scriptsize 54}$,
\AtlasOrcid[0000-0001-5763-2760]{N.J.~Abicht}$^\textrm{\scriptsize 48}$,
\AtlasOrcid[0000-0002-8496-9294]{S.H.~Abidi}$^\textrm{\scriptsize 30}$,
\AtlasOrcid[0009-0003-6578-220X]{M.~Aboelela}$^\textrm{\scriptsize 44}$,
\AtlasOrcid[0000-0002-9987-2292]{A.~Aboulhorma}$^\textrm{\scriptsize 36e}$,
\AtlasOrcid[0000-0001-5329-6640]{H.~Abramowicz}$^\textrm{\scriptsize 154}$
% ...
\bigskip
\\

% ...
$^{17}$Department for Physics and Technology, University of Bergen, Bergen; Norway.\\
% ...
$^{30}$Physics Department, Brookhaven National Laboratory, Upton NY; United States of America.\\
% ...
$^{36}$$^{(a)}$Facult\'e des Sciences Ain Chock, Universit\'e Hassan II de Casablanca;$^{(b)}$Facult\'{e} des Sciences, Universit\'{e} Ibn-Tofail, K\'{e}nitra;$^{(c)}$Facult\'e des Sciences Semlalia, Universit\'e Cadi Ayyad, LPHEA-Marrakech;$^{(d)}$LPMR, Facult\'e des Sciences, Universit\'e Mohamed Premier, Oujda;$^{(e)}$Facult\'e des sciences, Universit\'e Mohammed V, Rabat;$^{(f)}$Institute of Applied Physics, Mohammed VI Polytechnic University, Ben Guerir; Morocco.\\
% ...
$^{44}$Physics Department, Southern Methodist University, Dallas TX; United States of America.\\
% ...
$^{48}$Fakult\"{a}t Physik, Technische Universit{\"a}t Dortmund, Dortmund; Germany.\\
% ...
$^{54}$II. Physikalisches Institut, Georg-August-Universit\"{a}t G\"ottingen, G\"ottingen; Germany.\\
% ...
$^{83}$$^{(a)}$Khalifa University of Science and Technology, Abu Dhabi;$^{(b)}$New York University Abu Dhabi, Abu Dhabi;$^{(c)}$United Arab Emirates University, Al Ain;$^{(d)}$University of Sharjah, Sharjah; United Arab Emirates.\\
% ...
$^{102}$CPPM, Aix-Marseille Universit\'e, CNRS/IN2P3, Marseille; France.\\
% ...
$^{121}$Homer L. Dodge Department of Physics and Astronomy, University of Oklahoma, Norman OK; United States of America.\\
% ...
$^{154}$Raymond and Beverly Sackler School of Physics and Astronomy, Tel Aviv University, Tel Aviv; Israel.\\
% ...
$^{a}$ Also at Affiliated with an institute formerly covered by a cooperation agreement with CERN.\\
$^{b}$ Also at An-Najah National University, Nablus; Palestine.\\
$^{c}$ Also at Borough of Manhattan Community College, City University of New York, New York NY; United States of America.\\
$^{d}$ Also at Center for Interdisciplinary Research and Innovation (CIRI-AUTH), Thessaloniki; Greece.\\
$^{e}$ Also at Centre of Physics of the Universities of Minho and Porto (CF-UM-UP); Portugal.\\
$^{f}$ Also at CERN, Geneva; Switzerland.\\
% ...

\end{flushleft}


%

%
%

%

%
%
%
%

\end{document}